\documentclass[%
 reprint,
superscriptaddress,
 amsmath,amssymb,
 aps,
 pre
]{revtex4-2}

\usepackage[color=blue!20]{todonotes}

\usepackage{mathtools}
\usepackage{amsfonts}  
\usepackage[
  breaklinks=true
]{hyperref}
\hypersetup{
	anchorcolor = black,
	colorlinks=true,
	linkcolor=blue,
	filecolor=blue,      
	urlcolor=blue,
	menucolor =blue,
	pdfpagemode=FullScreen,
	breaklinks=true,
	citecolor=blue,
}

\usepackage[capitalise,nameinlink]{cleveref}
\crefname{section}{Sec.}{Secs.} 
\usepackage{graphicx}
\usepackage[caption=false]{subfig} 

\usepackage{color}
\usepackage{verbatim} 

\usepackage{natbib}
\usepackage[normalem]{ulem}

\begin{document}

\title{Noise-cancellation algorithm for simulations of Brownian particles}

\author{Regina Rusch}
\affiliation{Institut f\"ur Theoretische Physik, Technikerstra{\ss}e 21-A, Universit\"at Innsbruck, A-6020 Innsbruck, Austria.}
\author{Thomas Franosch}
\affiliation{Institut f\"ur Theoretische Physik, Technikerstra{\ss}e 21-A, Universit\"at Innsbruck, A-6020 Innsbruck, Austria.}
\author{Gerhard Jung}
\affiliation{Laboratoire Charles Coulomb (L2C), Universit\'e de Montpellier, CNRS, 34095 Montpellier, France.}
\date{\today}

\begin{abstract}
We investigate the usage of a recently introduced noise-cancellation algorithm for Brownian simulations  to enhance the precision of measuring transport properties such as the mean-square displacement or the velocity-autocorrelation function.
 The algorithm is based on explicitly storing the pseudo-random numbers  used to create the randomized displacements in computer simulations and subtracting them from the simulated trajectories. The resulting correlation function of the reduced motion is connected to the target correlation function up to a cross-correlation term. Using analytical theory and computer simulations, we demonstrate that the cross-correlation term can be neglected in all three systems studied in this paper. We further expand the algorithm to Monte Carlo simulations and analyze the performance of the algorithm and rationalize that it works particularly well for unbounded, weakly interacting systems in which the precision of the mean-square displacement can be improved by orders of magnitude. 

\end{abstract}

\maketitle

\section{Introduction}

Since Einstein’s  work in 1905 on the motion of suspended particles in fluids~\cite{Einstein_1905}, over 100 years of research have been devoted to better understand Brownian motion~\cite{Hanggi_2005,Xin_2016} with implications for statistical physics, biophysics~\cite{Frey_2005,Bressloff_2013}, biochemistry~\cite{Enkavi_2019}, and beyond.
Scientific progress has been achieved  through experiments~\cite{Saxton_1997} as well as the development of advanced theories~\cite{Franosch_2011}. 
Later, computer simulations have become essential for studying Brownian particles~\cite{Xavier_2010}. Computer simulations have matured into an important tool for understanding the complex behavior and transport of Brownian particles and making quantitative comparisons with theory and experiments~\cite{Alder_1967, Pancorbo_2017, Zembrzycki_2023, Qi_2017}.
The main limitation of computer simulations is the available computational resources, which restrict the precision of the extracted transport properties, such as the calculation of the mean-square displacement (MSD) or the velocity-autocorrelation function (VACF).
Therefore, extensive research has  been devoted to address these challenges, resulting in numerous algorithmic improvements and enhanced sampling methods suitable for specific contexts~\cite{Dubbeldam_2009,Yang_2019,Akhmatskaya_2008,Bernardi_2015,Henin_2022}.

In a recent study by Mandal \emph{et al.}~\cite{Mandal_2019}, a novel noise-cancellation (NC) algorithm was introduced to further enhance precision in Brownian dynamics simulations beyond the usage of suitable sampling techniques. The algorithm, motivated by an earlier algorithm by Frenkel~\cite{Frenkel_1987} for hopping transport on lattices, was used to investigate the VACF in dilute hard-sphere fluids and revealed a long-time tail in the VACF with a decay rate of $t^{-5/2}$, indicating the existence of persistent anti-correlations despite the absence of momentum conservation. These tails arise due to particle conservation and repeated encounters with scatterers and were predicted using analytical theory in the dilute regime 
\cite{Hanna_1981,Ackerson_1982,Felderhof_1983,Lowe_1995}.
Extracting this tail from the VACF, which is dominated by Brownian noise, is notoriously difficult and would require extensive computer simulations without the usage of the NC algorithm.

The basic idea of the algorithm is to store the pseudo-random numbers used to create the randomized displacements in computer simulations and subtract them from a simulated trajectory to obtain a reduced trajectory. Similar concepts can be found in earlier publications~\cite{Oettinger_1994, Melchior_1996, Evans_2008}. The MSD of the reduced trajectory contains the potential- and collision-induced displacements and can subsequently be linked to the standard MSD with an additional cross-correlation (CC) term. In Ref.~\cite{Mandal_2019} this CC term has been found to be negligible compared to the reduced MSD for the specific case of Brownian dynamics simulations of hard-sphere fluids.

In this paper we provide analytical support for the empirical findings of Ref.~\cite{Mandal_2019} since we see applications of the algorithm far beyond the specific system analyzed in Ref.~\cite{Mandal_2019}. We further supply detailed algorithms and guidelines for implementations, in particular for Monte Carlo simulations. Our strategy to increase the understanding of the algorithm and its limitations is to use quite simplistic systems which still allow for generalization of the results. 

We consider Brownian dynamics simulations~\cite{Mandal_2019}, and Monte Carlo simulations and apply the algorithm to three different one-dimensional potentials: the harmonic potential, the periodic barrier potential~\cite{Brinkman_1956,Moersch_1979}, and the cosine potential. Harmonic potentials are widely used in Brownian motion research, such as in optical tweezers, because of their simplicity and ability to confine particles~\cite{Zembrzycki_2023,Pancorbo_2017}. Periodic cosine potentials also have broad applications, including modulated liquids~\cite{Capellmann_2018}, fluctuations of Josephson supercurrents through a tunneling junction~\cite{Ambegaokar_1969}, lattice systems~\cite{Weiner_1974}, and directed transport~\cite{Sanchez-Palencia_2004,Jack_2022}.

Varying the strength of the external potentials enables us to study the performance of the NC algorithm. The performance is determined by comparing the error of the MSD obtained from the NC algorithm with the error of the MSD determined using standard simulations. We find a decrease of up to two orders of magnitude in the periodic models for small potential heights. In contrast, we found an increase of the error in systems in which the particle is bounded. In addition, we compute the VACF and illustrate the system dynamics, and confirm the algebraic  initial decay of $t^{-1/2}$ for the periodic potential in~\cref{sec:periodicbarriermodel}~\cite{Franosch_2010, Ackerson_1982,Hanna_1981}. 

The paper is organized as follows. The NC algorithm is introduced in~\cref{sec:noise_suppression_algorithm} and the implementation explained in~\cref{sec:implementation}. 
For the harmonic  model in~\cref{sec:toymodel} we solve analytically the discretized overdamped Langevin equation to show that the CC term vanishes for sufficiently short simulation time steps and compare the results to Brownian dynamics simulations. This also gives detailed insight into the non-trivial convergence behavior of the variance of this distribution. For the periodic  models in~\cref{sec:periodicbarriermodel} and~\cref{sec:cosine_potential} we perform Monte Carlo simulations to show that the CC term is much smaller than Brownian noise and analyze the performance of the model. Finally, a summary and conclusion are provided, and potential future applications of the NC algorithm are presented in~\cref{sec:summary}. 

\section{Noise-suppression algorithm} \label{sec:noise_suppression_algorithm}
\begin{figure}[tbp] 
	\centering
	\includegraphics[width=\linewidth]{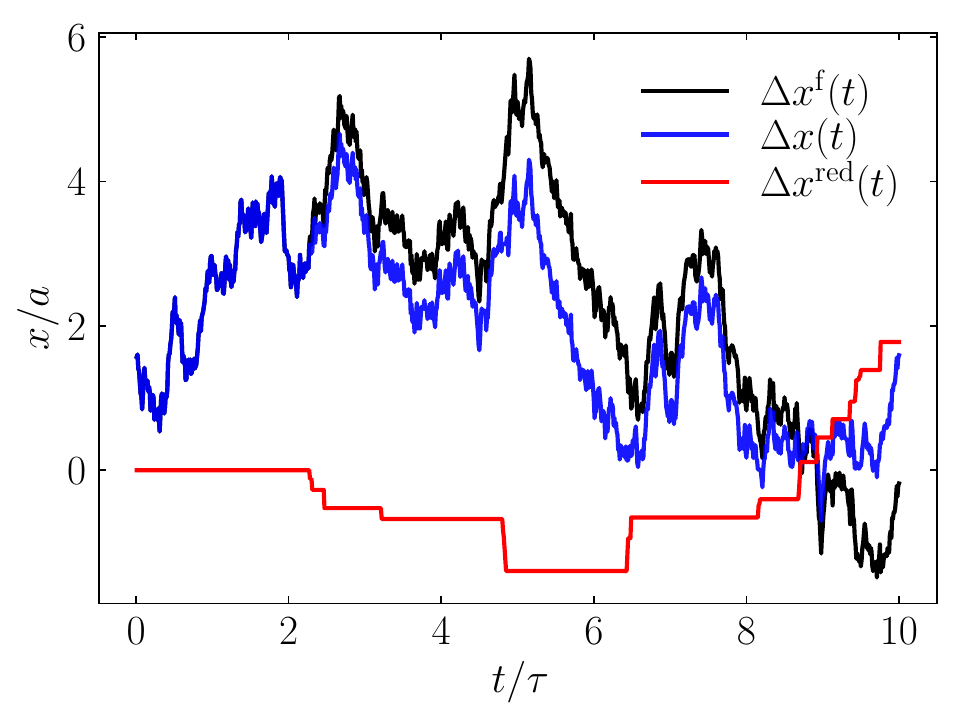}
	\caption{Trajectory of a free Brownian particle $\Delta x^\text{f}(t)$, a Brownian particle in a periodic barrier potential $\Delta x(t)$ with identical noise (see~\cref{sec:periodicbarriermodel}), and the  difference between the two, the reduced trajectory $\Delta x^\mathrm{red}(t)$. 
	}
	\label{fig:t4_NC_trajectories_step}
\end{figure}

The NC algorithm, motivated by Frenkel~\cite{Frenkel_1987} and elaborated by Mandal \emph{et al.}~\cite{Mandal_2019}, was implemented to increase the precision in computing the MSD $ \langle\Delta x(t)^{2}\rangle$ and its derived quantities, computed from particle displacements $\Delta x(t) = x(t) - x(0)$ calculated using standard simulations.
We define the VACF for a colloidal suspension 
 \begin{equation} \label{ref:VACF}
Z(t) \coloneq \frac{1}{2} \frac{d^2}{dt^2} \langle\Delta x(t)^{2}\rangle ,
 \end{equation}
such that it corresponds to the standard VACF in the Newtonian case.
The fundamental idea behind the algorithm is to generate an additional trajectory of free particles with \textit{identical} noise (i.e. pseudo-random numbers). The free particle displacement $\Delta x^\text{f}(t)  
= x^{\mathrm{f}}(t) - x^{\mathrm{f}}(0)$ can be subtracted from the original trajectory to obtain the reduced displacement 
\begin{align}
		\Delta x^\mathrm{red}(t) \coloneq \Delta x(t) - \Delta x^\mathrm{f}(t).
\end{align}
A visualization of the algorithm is displayed in~\cref{fig:t4_NC_trajectories_step} for the case of a one-dimensional periodic barrier potential. The free particle moves only due to Brownian noise. If this noise is subtracted from the trajectory, what remains is the reduced trajectory, which varies only due to interactions with the potential-energy landscape. 

The core concept of the NC algorithm involves rewriting the standard MSD in terms of the reduced MSD and the free particle MSD
\begin{align} \label{eq:NC_allterms}
	\begin{aligned}
		\langle \Delta x(t)^2 \rangle &=
		\langle \Delta x^\text{f}(t)^2 \rangle 
		-\langle\Delta x^\mathrm{red}(t)^{2}\rangle
		+2\langle\Delta x(t)  \Delta x^\mathrm{red}(t)\rangle .
	\end{aligned}
\end{align}

Importantly, the MSD for a free particle is noiseless, because it can be replaced with the exact result $\left\langle \Delta x^\mathrm{f}(t)^2 \right\rangle=2Dt$ in one dimension, where $D$ denotes the bare diffusion constant.
In addition, the CC term $\langle\Delta x(t) \Delta x^\mathrm{red}(t)\rangle$  will be shown to be often much smaller than the MSD of a free particle minus the reduced MSD. If the CC term can be neglected, the MSD can therefore be computed by subtracting the reduced MSD from the solution of a free particle
\begin{align} \label{ref:NC_MSD}
		\langle\Delta x(t)^{2}\rangle& \approx 2 D t-\langle\Delta x^\mathrm{red}(t)^{2}\rangle . 
\end{align}
This is the core relation for the NC algorithm as it shows that the calculation of the full MSD can be replaced by an exact result combined with a - hopefully - much less noisy interaction term $\langle\Delta x^\mathrm{red}(t)^{2}\rangle$.
It should also be mentioned that neglecting the CC term simplifies the implementation and eliminates the component containing Brownian noise. However, if the CC term is not negligible but adds sufficiently less noise, the algorithm implemented as in~\cref{eq:NC_allterms} is still exact and can perform well.

Inserting~\cref{ref:NC_MSD} into the definition~\cref{ref:VACF}, we similarly find an expression for the VACF
\begin{align} \label{ref:NC_VACF}
	Z(t) = \frac{1}{2} \frac{d^2}{dt^2} \langle\Delta x(t)^{2}\rangle \approx - \frac{1}{2}  \frac{d^2}{d t^2}  \langle\Delta x^\mathrm{red}(t)^{2}\rangle= Z_\text{NC}(t).
\end{align}
This relation shows that the VACF of the interacting particle can be determined from the reduced trajectories, thereby increasing the precision of the simulation by typically two orders of magnitude. Therefore, the algorithm is particularly suitable for unraveling subtle correlations hidden beneath the dominant Brownian noise. 

\section{Implementation of the NC algorithm} \label{sec:implementation}
The algorithm can be applied to Brownian dynamics, 
\begin{align}\label{eq:Brownian_eom}
\frac{d}{d t} x(t) = \mu F(x(t)) + \eta(t),
\end{align}
which describes the evolution of the particle position $x(t)$ at time $t$. Here $F(x(t))$ is the force acting on the particle and  $\mu$ denotes the mobility. For simplicity, we restrict the discussion to one-dimensional systems, however, the considerations can be readily transferred  to higher dimensions or interacting particles.
The Brownian particle in the overdamped regime (no inertia) moves due to thermal fluctuations modeled by Gaussian white noise $\eta (t)$ with the properties
\begin{align} \label{eq:noise_properties}
	\left\langle \eta(t) \right\rangle =0 , \qquad	\left\langle \eta(t)\eta(t^{\prime}) \right\rangle = 2 D \delta(t-t^{\prime}),
\end{align} 
 where $D$ is the (short-time) diffusion coefficient.  Temperature enters the problem via the Einstein relation $D= \mu k_\text{B} T$.  

Similarly, the NC algorithm can be used in combination with Monte Carlo simulations in which the free trajectory without interactions uses the \emph{identical} trial moves in each Monte Carlo step as the standard simulation with interactions.

To apply this algorithm in computer simulations, we generate two trajectories relying on the \emph{identical} noise history. We discretize the  equation of motion,~\cref{eq:Brownian_eom} with a time step $\Delta t$ such that $x_n $ approximates $x(n \Delta t)$.  Using the Euler-Maruyama discretization~\cite{Burrage_2000, Maruyama_1955, Ermak_1978} we find the iteration scheme
\begin{subequations}
\begin{align}
x_{n+1} &= x_n + \mu F(x_n) \Delta t + \eta_{n},  \\
x_{n+1}^{\mathrm{f}} &= x_n^{\mathrm{f}} + \eta_{n},  \label{eq:discrete_free}
\end{align}
\end{subequations}
with $n \in \mathbb{Z}$. Here $\eta_n$ are Gaussian random variables characterized by
\begin{align} \label{eq:noiseproperiesdiscrete}
\langle \eta_n \rangle &= 0, \qquad \langle \eta_n \eta_m \rangle = 2 D \Delta t \delta_{nm}. 
\end{align}
Rather than calculating the free motion,~\cref{eq:discrete_free}, we eliminate the free dynamics in favor of  the reduced motion $ x_n^{\mathrm{red}} \coloneq x_n - x_n^{\mathrm{f}}$ with iteration scheme
\begin{subequations} \label{eq:discrete_red}
\begin{align}
x_{n+1} &= x_n + \mu F(x_n) \Delta t + \eta_{n},  \\
x_{n+1}^{\mathrm{red}} & =  x_n^{\mathrm{red}} + \mu F(x_n) \Delta t,
\end{align}
\end{subequations}
highlighting that the reduced dynamics is much smoother than the original one. Ensemble-averaged observables such as the MSD are then obtained as moving time averages of the corresponding discretized version. 
The standard MSD can be derived from the reduced MSD, as in~\cref{ref:NC_MSD}. The VACF of the system can be approximated by the negative discrete second derivative of the reduced MSD as indicated in~\cref{ref:NC_VACF}.

Importantly, the implementation and computational effort of the algorithm are minimal, as they only require calculating the difference to a second trajectory of free particles and no additional force calculations are necessary, which are the bottleneck of most computer simulations.
	
\section{Brownian particle in a harmonic potential} \label{sec:toymodel} 
In this section, we demonstrate that the CC term is negligible in the harmonic relaxator and thus~\cref{ref:NC_VACF,ref:NC_MSD} are valid. We calculate the CC term analytically by solving the discretized overdamped Langevin equation and compare Brownian dynamics simulations to the analytic solution. 
The model consists of a one-dimensional, single Brownian particle in a harmonic potential
\begin{align}
	U(x)= \frac{k}{2} x^2,
\end{align}
with spring constant $k$. Then, the discretized equation of motion~\cref{eq:discrete_red} reads
 \begin{subequations}\label{eq:discrete_red_harmonic}
\begin{align}
x_{n+1} &= x_n - x_n  \Delta t/\tau + \eta_{n},  \\
x_{n+1}^{\mathrm{red}} & =  x_n^{\mathrm{red}} - x_n\Delta t/\tau,
\end{align}
\end{subequations}
with the trap relaxation time $\tau = 1/\mu k$.  By induction one finds the closed solution
\begin{subequations}\label{eq:discrete_red_harmonic_induction}
\begin{align}
x_n &= x_0 r^n + \sum_{k=1}^n r^{n-k} \eta_{k}, \\
x_n^{\text{red}} &= x_0 (r^n-1) + \sum_{k=1}^n (r^{n-k}-1) \eta_{k} ,
\end{align}
\end{subequations}
with the abbreviation $r= 1- \Delta t/\tau$ where $\Delta t \ll \tau$ is anticipated. 
\begin{figure}[tbp]
	\centering
	\subfloat{
		\includegraphics[width=\linewidth]{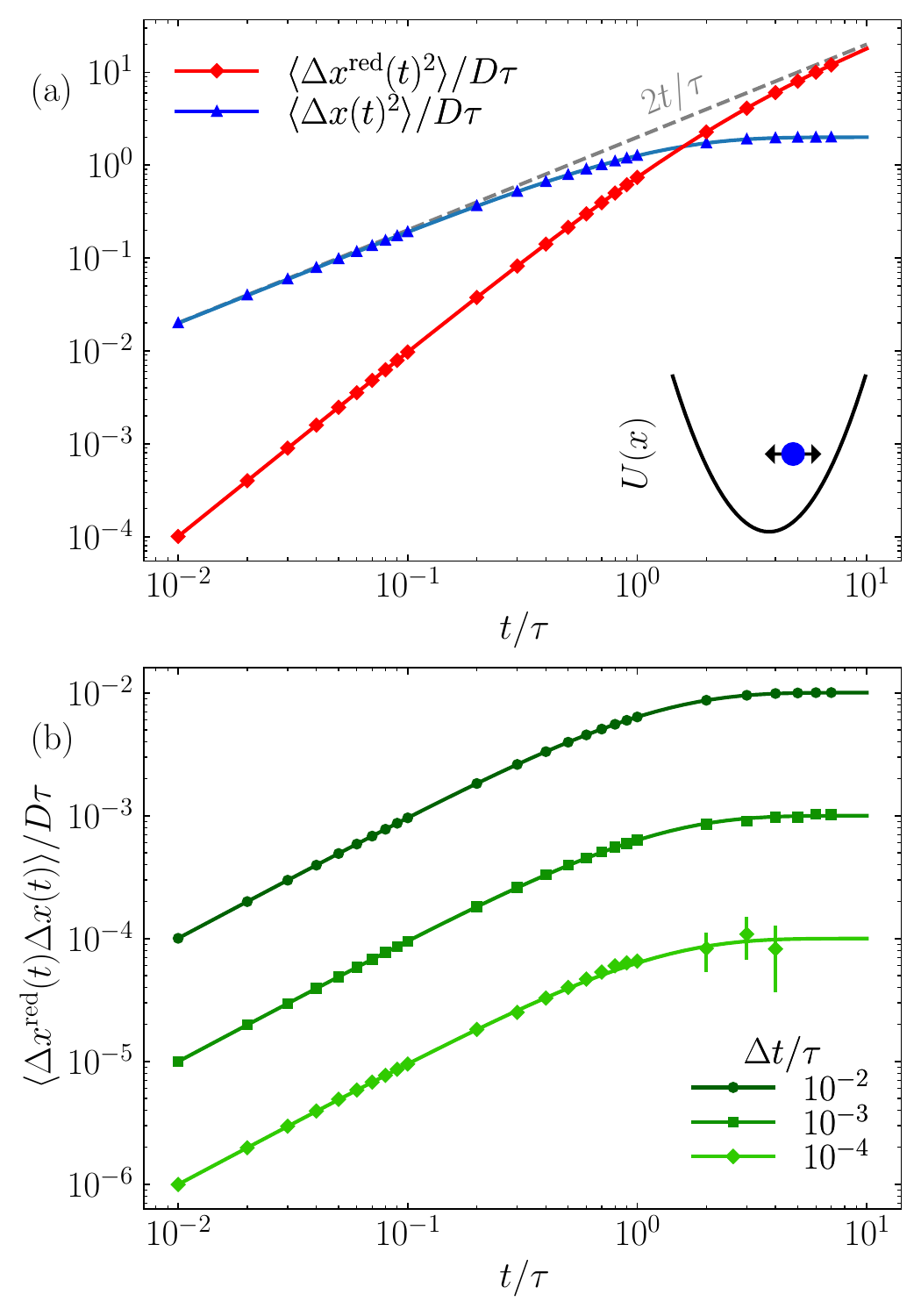}
	  	\label{fig:t2_HO_MSDS}}
	\subfloat{
	  \includegraphics[scale=0.0]{Fig_2.pdf}
	  \label{fig:HP_CC_term}
	}
	\caption{ 
	Analytical results in lines and simulation results in symbols for each term in~\cref{eq:NC_allterms} for the harmonic relaxator.
	(a) Reduced MSD $\left\langle \Delta x^\mathrm{red}(t)^2\right\rangle$ and standard MSD $\left\langle \Delta x(t)^2\right\rangle$ for $\Delta t/\tau=0.01$. The dashed gray line represents the free diffusion  and serves as a guide for the eye.
	(b) CC term $\langle \Delta x^\mathrm{red}(t) \Delta x(t) \rangle $ for different simulation time steps $\Delta t$.  
	}
	
	\label{fig:t2_HO_MSDS_CC} 
  \end{figure}
The angular brackets are to be understood as ensemble averages which in simulations, are replaced by time averages. 
We therefore consider the simulation to have started in the remote past such that the system is in equilibrium at  time zero.  
Correspondingly  $x_0$ is drawn from the stationary probability distribution of  the discrete stochastic process $(x_n)_{n\in \mathbb{Z}}$. From \cref{eq:discrete_red_harmonic} one can see that for a stationary process, it holds that 
\begin{align}
	\langle x_{n+1} \rangle = r \langle x_n \rangle + \langle \eta_n \rangle  \overset{!}{=} \langle x_n \rangle,
\end{align}
which implies $\langle x_n \rangle = 0 $ as $0<r<1 $. We find that $x_n$ is a centered Gaussian variable  and for the variance in the stationary case we impose 
 \begin{align}
	\text{Var}[x_{n+1}]=r^{2} \text{Var}[x_{n}] +\text{Var}[\eta _n] \overset{!}{=}\text{Var}[x_{n}] ,
\end{align}
from which 
\begin{align}
	\text{Var}[x_{n}]=\frac{2D \Delta t}{1-r^2}= D \tau +\frac{D \Delta t}{2}+O(\Delta t^2),
\end{align}
follows.  This shows that the discrete Langevin equation does not lead exactly to the Boltzmann distribution with variance $D \tau = k_\text{B} T /k$ but includes  a correction term vanishing only in the limit $\Delta t\to 0$.
The CC term is then readily calculated from \cref{eq:discrete_red_harmonic_induction} since after averaging over the thermal noise, only contributions from the initial condition and diagonal terms of the noise survive. We find 
\begin{align}   \label{eq:cross-corr-discrete-1}
	\langle \Delta x_n^\mathrm{red} \Delta x_n \rangle = \langle x_0^2  \rangle ( 1-r^n)^2 + 2 D \Delta t \sum_{k=1} (r^{n-k}-1) r^{n-k}  ,
\end{align}
where $\Delta x_n= x_n -x_0$ and  $\Delta x_n^\textrm{red}= \Delta x_n-\Delta x_n^\textrm{f} = x_n^\textrm{red}$ since at time zero the positions of the free and interacting particle coincide. 
From the derived distribution, we find that $\langle x_0^2 \rangle= 2D \Delta t/(1-r^2) $ and expression \cref{eq:cross-corr-discrete-1} simplifies to 
\begin{align} \label{eq:cross-corr-discrete}
	\left\langle \Delta x^\mathrm{red}_n \Delta x_{n} \right\rangle&=\frac{2 D  \Delta t ( 1-r^n)}{1+r} \nonumber \\ 
	&= D \text{$\Delta $t} \left(1-e^{-t/\tau }\right)+O(\Delta t^2),
\end{align}
where in the second line we anticipate the continuum limit $\Delta t \to 0, n\to \infty$ with $t = n \Delta t$  fixed. Reinstating continuous time, we find that the CC term
\begin{align}
\langle \Delta x^\mathrm{red}(t) \Delta x(t) \rangle = 0,
\end{align}
vanishes for the harmonic relaxator and~\cref{eq:cross-corr-discrete} reveals that convergence is linear in the time step $\Delta t$. 
For reference, we also provide the expressions for the MSDs for the discretized form
\begin{subequations}\label{eq:msds-corr-discrete}
\begin{align}
	\langle \Delta x(t)^2 \rangle &=\frac{4 D \Delta t \left(1-r^n\right)}{1-r^2}  \nonumber \\
	&= 2 D \tau (1-e^{t/\tau})  + O(\Delta t), \\
	\langle \Delta x^\mathrm{red}(t)^2 \rangle &= 2 D \Delta t \left(n-\frac{2 r \left(1-r^n\right)}{1-r^2}\right) \nonumber \\ 
	&= 2 D t -2 D \tau (1-e^{t/\tau}) + O(\Delta t).
\end{align}
\end{subequations}

We compare Brownian dynamics simulations to analytical results using the discretized overdamped Langevin equation,~\cref{eq:discrete_red_harmonic}.  To save simulation time and memory, we use the order-$n$ algorithm described in  Frenkel and Smit~\cite{Frenkel_2002} to calculate the MSD on a logarithmic scale for all simulations performed in this paper. As expected, the CC term decreases as the simulation time step $\Delta t$ is decreased, see~\cref{fig:HP_CC_term}. The simulated data are in very good agreement with the analytical solution found in~\cref{eq:cross-corr-discrete,eq:msds-corr-discrete}. The CC term increases linearly in time for short times $t \ll \tau$ and reaches a constant value, depending  on the simulation time step $\Delta t$ for $t \gg \tau$. The CC term in \cref{eq:NC_allterms} is negligible, as it is orders of magnitude smaller than the other terms of the equation, which are the difference between the free particle and the reduced MSD.
\begin{figure}[tbp] 
	\centering
	\includegraphics[width=\linewidth]{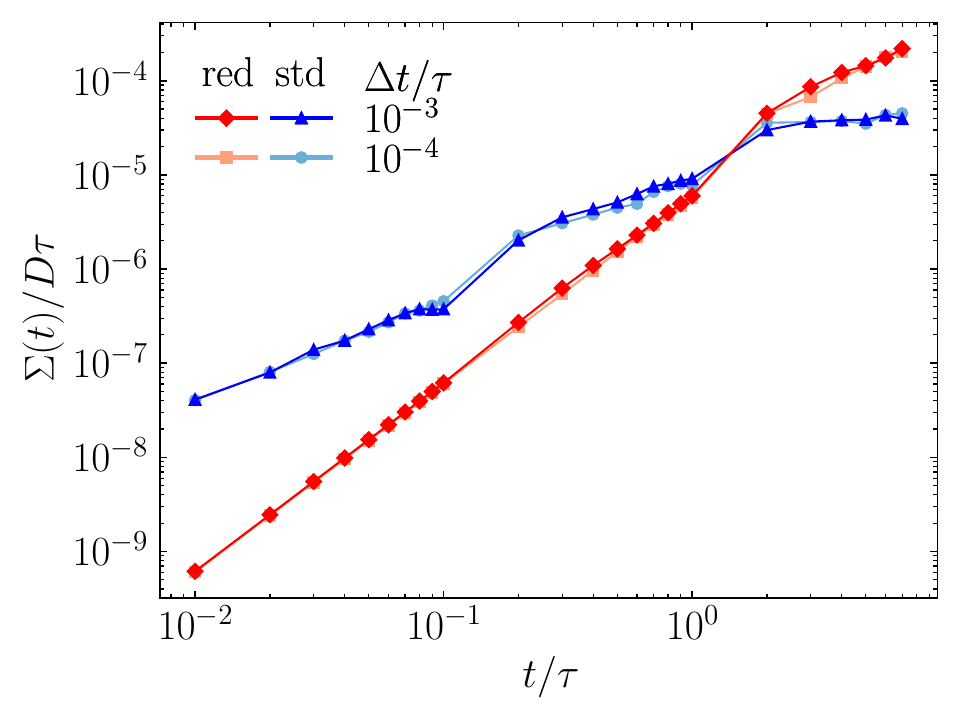}
	\caption{Comparison of the standard error of the mean $\Sigma (t)$ of the reduced MSD computed using the NC algorithm and the standard (STD) MSD for different simulation time steps for a harmonic relaxator.}
	\label{fig:MSD_std_HP}
\end{figure}

Besides the analysis of the CC term we investigated the performance of the NC algorithm applied to the harmonic relaxator.  We compare the error of the reduced MSD obtained using the NC algorithm with the error of the MSD obtained using the standard algorithm. To quantify the error, we use the standard error of the mean (SEM), which is extracted from $n$ independent simulations. The SEM $\Sigma $ of the MSD is then computed as
\begin{subequations}\label{eq:stdMSD}
\begin{align} 
	\Sigma &= \sqrt{\frac{\sigma^2 }{n}} , \\
	\sigma^2 &= \frac{1}{n-1} \sum_{i=1}^n (\langle \Delta x(t)^2 \rangle_i-\overline{\langle \Delta x(t)^2 \rangle})^2, 
\end{align}
\end{subequations}
where $\sigma^2 $ is the sample variance and $\overline{\langle \Delta x(t)^2 \rangle}$ is the sample mean. We find that the SEM of the MSD is, for long times, larger when using the NC algorithm compared to standard simulations (see~\cref{fig:MSD_std_HP}) for both simulation time steps. At these times, the MSD displays a plateau due to the confining potential, while both the reduced MSD and the CC term increase linearly (see~\cref{fig:HP_CC_term}). We will further discuss the efficiency of the algorithm when applied to (temporarily) trapped particles in the next sections.

\section{Brownian particle in a periodic barrier potential} \label{sec:periodicbarriermodel}
The objective of this section is to determine how well the NC algorithm performs for different potential barriers. The performance is determined by comparing the precision of the reduced MSD from the NC algorithm to the MSD of the standard method. We apply the algorithm to a periodic barrier model, vary the potential height, and determine the ranges in which  the algorithm’s  performance is favorable.Analyzing the underlying factors contributing to these findings, we can draw conclusions that may be generalizable to other systems and indicate important applications of the method.

The system of choice is a single  Brownian particle in a one-dimensional periodic potential with period $a$. The potential consists of constant regions only and is defined by
\begin{align} \label{eq:steppotential}
	U(x) = \begin{cases}
		\Delta U & \text{for } - a/2 < x \leq 0, \\
		0 & \text{for } 0 < x \leq  a/2 ,
	\end{cases}
\end{align}
with potential height $\Delta U$ 
and continued periodically
\begin{align}
	U(x) = U(x+  a) .
\end{align}
Since $\Delta U>0$, there are jumps in the potential at $x=  m  a$ and $x=(m+1/2)  a$ for integer $m\in\mathbb{Z}$.  Then, the period of the potential $a$ sets the unit of length, while $\tau \coloneq a^2 /D$ is  the natural time scale of the problem. There is a single dimensionless control parameter corresponding to the reduced potential step $ \Delta U/k_\text{B} T$. 
Here  $D$ is the (bare) diffusion constant of the Brownian particle and $k_\text{B} T$ is the thermal energy.

\begin{figure}[tbp]
	\centering
	\subfloat{
		\includegraphics[width=\linewidth]{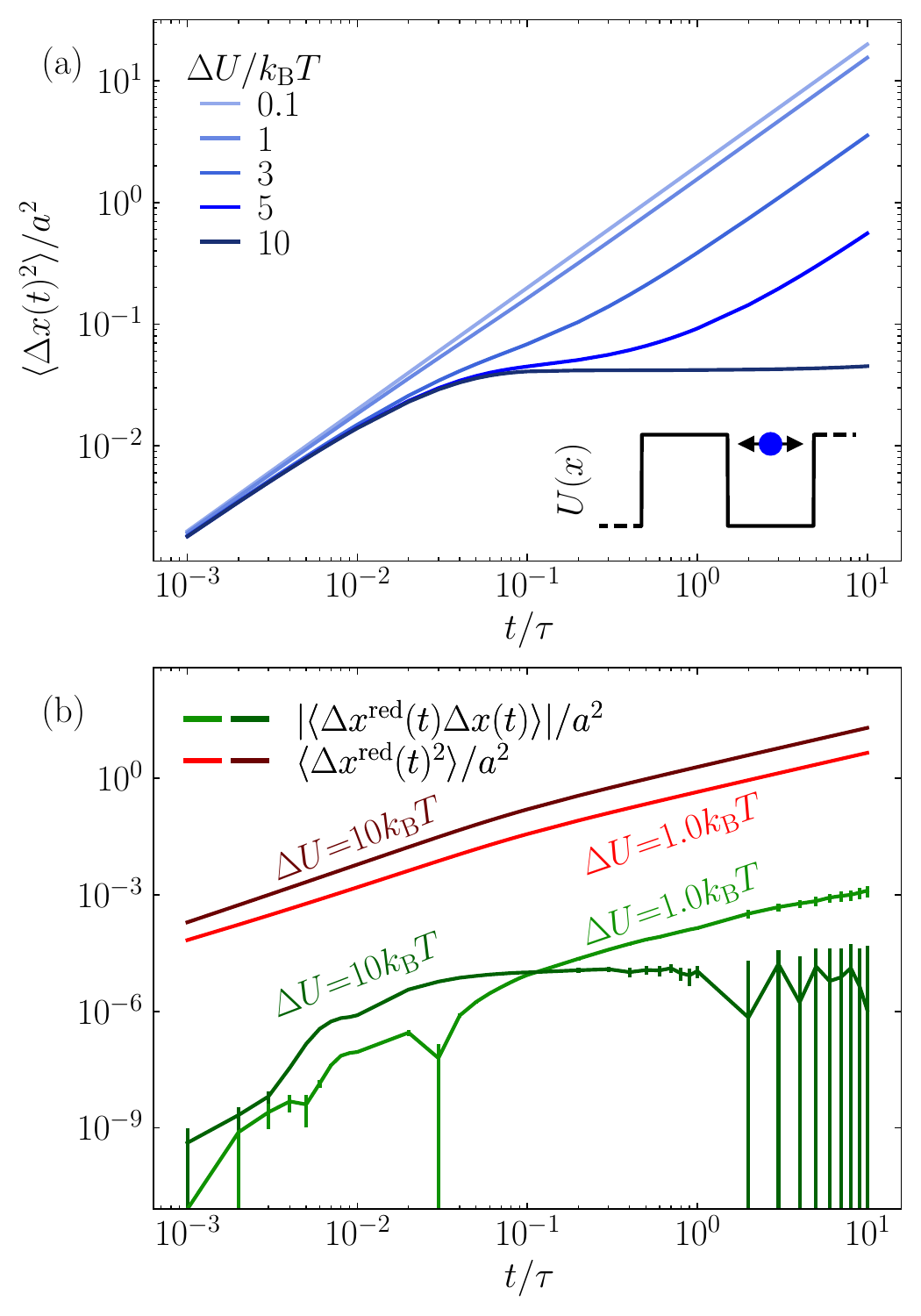}
	  	\label{fig:MSD_periodic_barrier}}
	\subfloat{
	  \includegraphics[scale=0.0]{Fig_4.pdf}
	  \label{fig:MSD_nc_cc_periodic_barrier2}
	}
	\caption{ 
	Simulation results for each term in~\cref{eq:NC_allterms} for a Brownian particle in a periodic barrier potential for different potential heights $\Delta U$.
	(a) Standard MSD $\left\langle \Delta x(t)^2\right\rangle$. 
	(b) Reduced MSD $\left\langle \Delta x^\mathrm{red}(t)^2\right\rangle$ and the absolute value of the CC term $ |\left\langle\Delta x(t)\Delta x^\mathrm{red}(t)\right\rangle |$. }
 	\label{fig:MSD_all_periodic_barrier}
  \end{figure}

We perform Monte Carlo simulations using the Metropolis algorithm to simulate the Brownian particle~\cite{Allen_2017,Metropolis_1953,Hansen_2007}. Therefore, we extend the NC algorithm in the following for the case of Monte Carlo simulations. We use discrete time steps $\Delta t$ and denote our approximant for $x(n \Delta t) $ by $x_n$.  At each time, a Gaussian random number with vanishing mean $\langle \eta_n\rangle $ and variance $\langle \eta_n^2 \rangle = 2 D \Delta t$ is used. The trial move for the trajectory of the problem is $x_{n+1}^*= x_n+\eta_n$. 
Then, we sample a binary random variable $X_n$ such that the probability for 
$X_n = 1$ (accept move)
is
\begin{align}
\text{Prob}(X_n=1) =  \text{min}\left( e^{-[ U(x_{n+1}^*) - U(x_n) ]/k_B T}, 1\right),
\end{align} 
 and $\text{Prob}(X_n=0) = 1-\text{Prob}(X_n=1)$ (reject move). 
 The new positions for the motion of the problem and the free motion are then
\begin{subequations}
\begin{align}
x_{n+1} &= x_n+ X_n \eta_{n} ,\\
x_{n+1}^\text{f}  &= x_n^\text{f} + \eta_{n}.
\end{align}
\end{subequations}
Note that for the free motion the trial move is always accepted. 
For the reduced motion we find the more compact expression
\begin{subequations}
\begin{align}
x_{n+1} &= x_n+ X_n \eta_{n} ,\\
x_{n+1}^\text{red}  &= x_n^\text{red} + (X_n-1) \eta_{n}.
\end{align}
\end{subequations}
The last relation highlights that the reduced motion changes only if the trial  move of the original trajectory is rejected, $X_n=0$. This can only happen if the particle attempts to climb the potential, which in this case of a stepwise potential can occur only at jumps.  
\begin{figure}[tbp] 
	\centering
	\includegraphics[width=\linewidth]{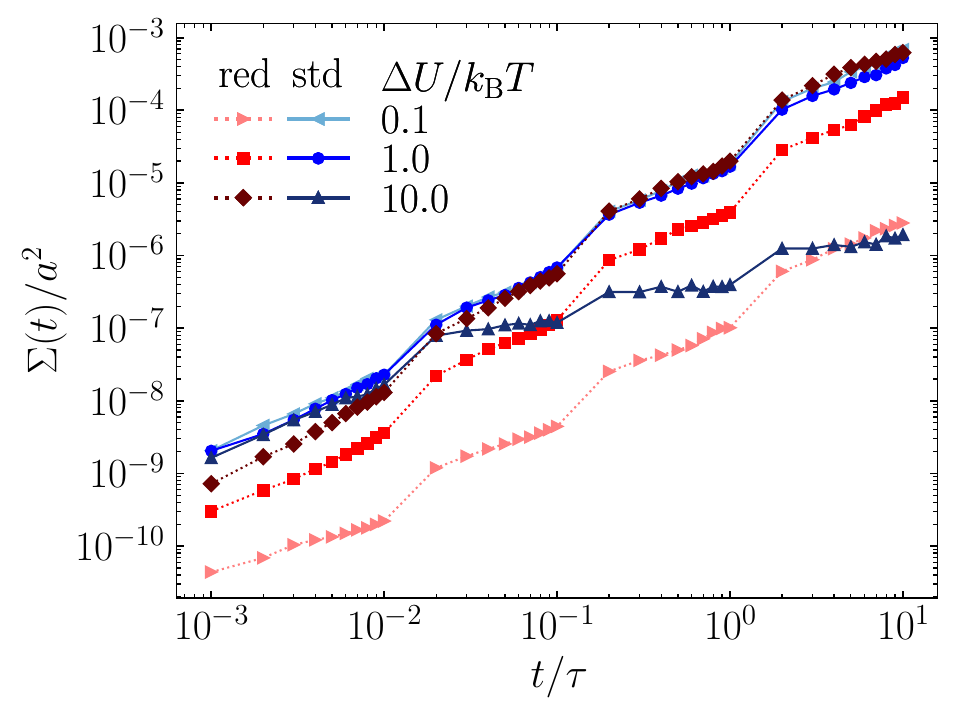}
	\caption{Comparison of the standard error of the mean $\Sigma (t)$  of the reduced MSD computed using the NC algorithm and standard MSD for different barrier heights for a Brownian particle in a periodic barrier potential.}
	\label{fig:MSD_std_periodic_barrier}
\end{figure}
We apply the NC algorithm to the model for different barrier heights $\Delta U /k_\text{B} T$.  At short times $t \ll \tau$, the MSD corresponds to  free diffusion $\langle \Delta x^\text{f}(t)^2 \rangle=2Dt$ (see~\cref{fig:MSD_periodic_barrier}).  For longer times, a plateau emerges for sufficiently high  barriers at time scales where the particle is temporarily  trapped  between the barriers ($t\simeq \tau$). The particle will not remain trapped indefinitely because  there is a non-zero probability that it will  overcome the barrier. Therefore, for a sufficiently long observation time $t \gg \tau$, each MSD grows linearly in time again with a long-time diffusion coefficient that decreases as the barrier height increases.

In~\cref{eq:NC_allterms} the standard MSD is decomposed into the reduced MSD $\langle \Delta x^{\text{red}}(t)^2 \rangle$, the CC term $\langle \Delta x^{\text{red}}(t) \Delta x(t) \rangle $  and the MSD of a free particle $\langle \Delta x^{\text{f}}(t)^2 \rangle$. In the following, these terms and their variances are analyzed. The CC term assumes positive and negative values. For better visualization, we show the absolute value of the CC term (see~\cref{fig:MSD_nc_cc_periodic_barrier2}). For the given potential heights, the absolute value of the CC term is several orders of magnitude smaller than the MSD of the free particle minus the reduced MSD and is thus negligible.
\begin{figure}[tbp]
	\centering
	\includegraphics[width=\linewidth]{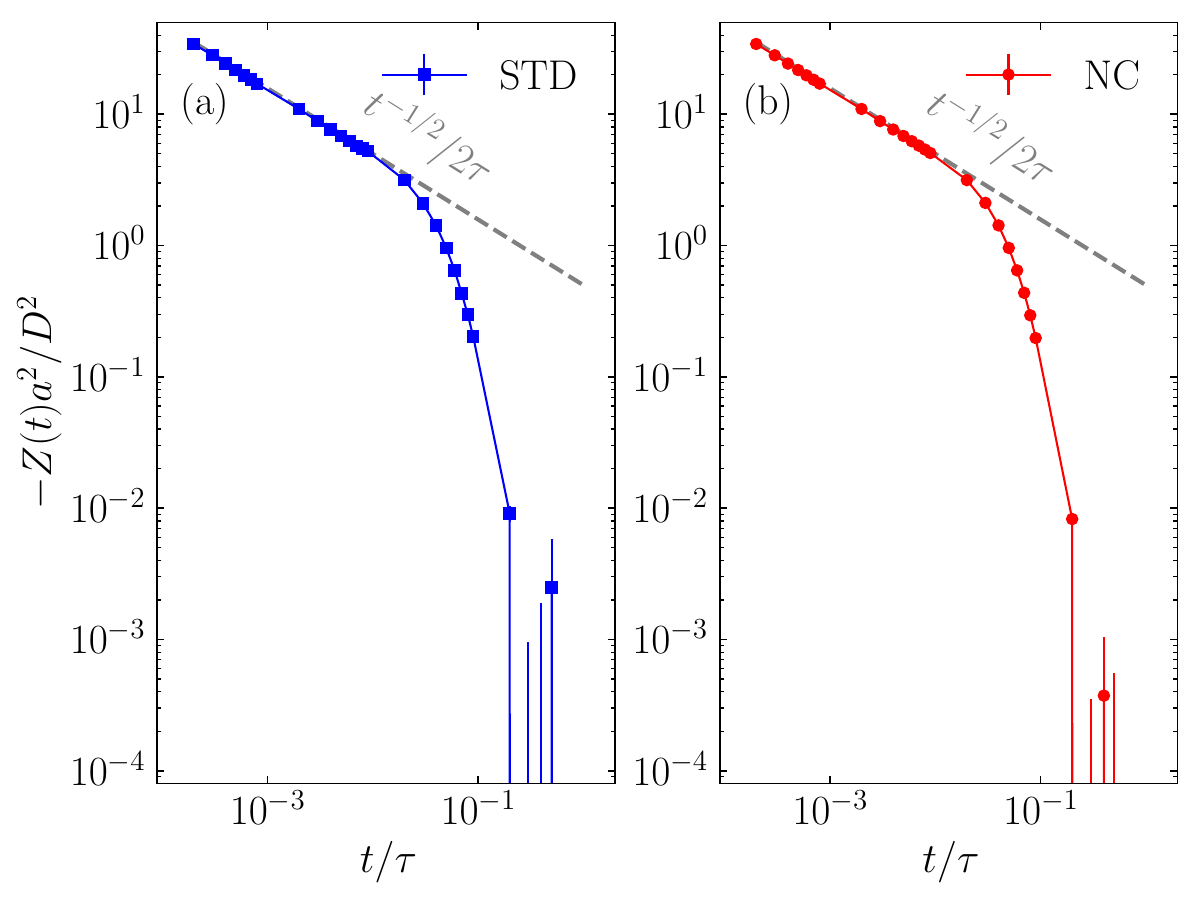}
	\caption{VACF for the periodic barrier potential computed  using the two different algorithms, standard and NC on double-logarithmic scales for $U=1.0 k_\text{B} T$. The dashed line serves as a guide for the eye. 
	}
	\label{fig:VACF_step}
\end{figure}

In general, similar to the standard MSD, the reduced MSD increases linearly with time for longer times. Yet, unlike the standard MSD displaying a plateau for high potential barriers at intermediate times $t \simeq  \tau$, the reduced MSD is not bounded for all times, because it always contains contributions from the unbounded motion of the free particle.

To analyze the performance of the NC algorithm we compare it to standard computer simulations, which correspond in this case to Monte Carlo simulations of the Brownian particle in a periodic barrier potential. More precisely, we  compare the SEM,~\cref{eq:stdMSD}, of the reduced MSD obtained using the NC algorithm to the SEM of the MSD obtained using the standard algorithm. We find that the NC algorithm reduces the SEM of the MSD by approximately one order of magnitude for barrier heights in the range of $ \Delta U \lesssim k_\text{B} T$, see~\cref{fig:MSD_std_periodic_barrier}. For a barrier height of $\Delta U = 0.1 k_\text{B} T$, the SEM is reduced by two orders of magnitude compared to the standard method. This can be intuitively understood because the dynamics in systems with small barriers is dominated by free diffusion, which is suppressed in the NC algorithm.

Consistent with the analysis of the harmonic relaxator, we find that the performance of the NC algorithm is poor for times $t \gtrsim  \tau$ when considering barrier heights of $\Delta U \gtrsim 10\,k_\text{B} T$,  for which the SEM for the reduced MSD exceeds that of the standard MSD. We rationalize this observation using the amplitude of the standard and the reduced MSD. For sufficiently high barriers, the standard MSD exhibits a plateau at intermediate times and is therefore smaller in amplitude than the reduced MSD, which grows  at least linearly. The smaller amplitude also leads to smaller variances as visible in~\cref{fig:MSD_std_periodic_barrier}. Consequently, the effect of the plateau in the MSD, which reduces the variance, is stronger than the error reduction achieved by the NC algorithm. Only for short times $t \ll \tau$, the performance is good because  the reduced trajectory is free of the dominant Brownian noise, resulting in higher precision compared to the standard trajectory.

Finally, we also calculate the VACF from the MSD using~\cref{ref:NC_VACF} for  $\Delta U = 1.0 k_\text{B} T$. The error of the VACF is computed with error propagation from the SEM of the MSD. The VACF is not following an exponential  for short times but exhibits a power-law decay of  $t^{-1/2}$ (see \cref{fig:VACF_step}).  This is due to the presence of barriers and reflection of the hard walls~\cite{Franosch_2010, Ackerson_1982,Hanna_1981}.  The VACF of the algorithm coincides with the standard one, however, the error is reduced by the algorithm. 
The algorithm excels in potentials where the diffusion is minimally suppressed, as indicated by $\langle \Delta x^{\text{red}}(t)^2\rangle \ll \langle \Delta x(t)^2\rangle $. In these cases the correction of the MSD to free diffusion is small. Nonetheless, this minor correction is important for computing the VACF, and the performance gain of the NC algorithm is significant.

\section{Brownian particle in a  cosine potential} \label{sec:cosine_potential}

We apply the same algorithm and analysis tools to a single Brownian particle in a periodic cosine potential and compare the results with the non-continuous periodic barrier potential discussed  in the previous~\cref{sec:periodicbarriermodel}. The purpose of this analysis is to investigate the dependance  of our conclusions on the details of the potential.

We again analyze all terms of~\cref{eq:NC_allterms} for different system parameters and analyze the performance of the algorithm by comparing the SEM,~\cref{eq:stdMSD}, of the reduced MSD obtained from the NC algorithm to the MSD obtained with the standard method.
The system consists of a single  Brownian particle in a one-dimensional cosine potential
\begin{align}
	U(x)=\frac{\Delta U}{2} \cos(2\pi x/a) .
\end{align}
We choose the period $a$  and the peak-to-peak amplitude $\Delta U$ of the cosine potential to be the same as for the periodic barrier potential.
  In contrast to the periodic barrier potential, the cosine potential is smooth and continuous with localized minima.

\begin{figure}[btp]
	\centering
	\subfloat{ 
		\includegraphics[width=\linewidth]{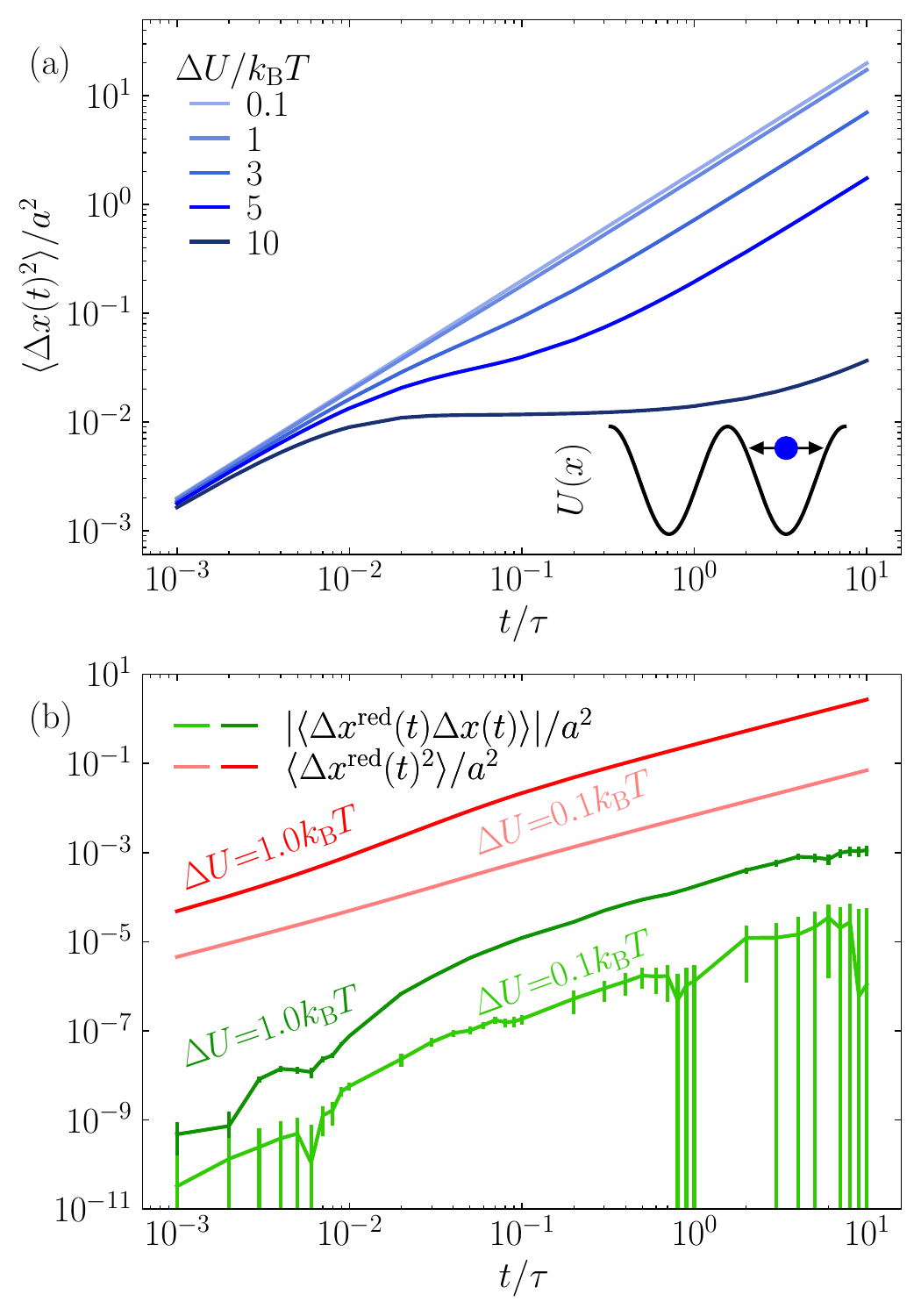} 
		\label{fig:MSD_sinus} }
		\subfloat{ 
		\includegraphics[scale=0.0]{Fig_7.pdf}
		\label{fig:MSD_nc_cc_sinus} }
	\caption{Simulation results for each term in~\cref{eq:NC_allterms} for a Brownian particle in a cosine potential for different peak-to-peak amplitudes $\Delta U $. 
	(a) Standard MSD $\left\langle \Delta x(t)^2\right\rangle$. 
	(b) Reduced MSD $\left\langle \Delta x^\mathrm{red}(t)^2\right\rangle$ and the absolute value of the CC term  $|\left\langle  \Delta x^\mathrm{red}(t)\Delta x(t)\right\rangle|$.  
	}
	\label{fig:MSD_all_sinus}
  \end{figure}

We apply the algorithm to a range of different peak-to-peak amplitudes $\Delta U$. We find that the qualitative behavior of the particle in the cosine potential is similar to that of the particle in the periodic barrier potential. For $\Delta U \lesssim k_\text{B} T$, the particle undergoes almost pure diffusion, whereas  for higher $\Delta U \gtrsim k_\text{B} T$ a plateau emerges in the MSD for intermediate times (see~\cref{fig:MSD_sinus}).

  \begin{figure}[tbp]
	\centering
	\includegraphics[width=\linewidth]{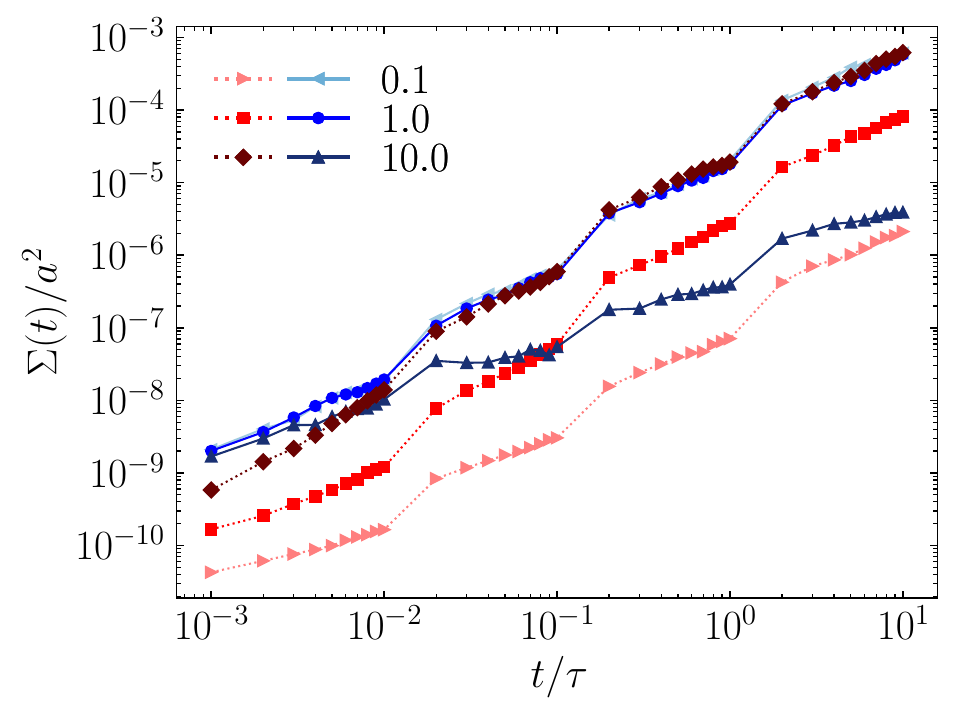}
	\caption{Comparison of the standard error of the mean $\Sigma (t)$ of the reduced MSD computed using the NC algorithm and standard MSD for different peak-to-peak amplitudes $\Delta U$ for a Brownian particle in a cosine potential.} 
	\label{fig:error_MSD_sinus} 
\end{figure}

Similar as described in~\cref{sec:periodicbarriermodel} we observe that the absolute value of the CC term is again several orders of magnitude smaller than the MSD of a free particle minus the reduced MSD and can therefore be neglected, see~\cref{fig:MSD_nc_cc_sinus}. Consistent with the previous results, we infer that the algorithm works better for systems with lower barriers as shown in~\cref{fig:error_MSD_sinus}. In fact, the improvement is even in quantitative agreement with our results of~\cref{sec:periodicbarriermodel}, as for the potential height of $\Delta U = 1.0 k_\text{B}T$ we observe an improvement of one order of magnitude, and for $\Delta U = 0.1\, k_\text{B}T$ the improvement is already two orders of magnitude. This is a  crucial finding: it shows that the details of the interactions are not relevant for the analysis and performance of the NC algorithm. Therefore, we can expect that the analysis performed and the conclusions drawn in the present paper are relevant for very general soft-matter systems and could therefore be used as a guide to which systems the NC algorithm should be applied to.

\begin{figure}[tbp]
	\centering
	\includegraphics[width=\linewidth]{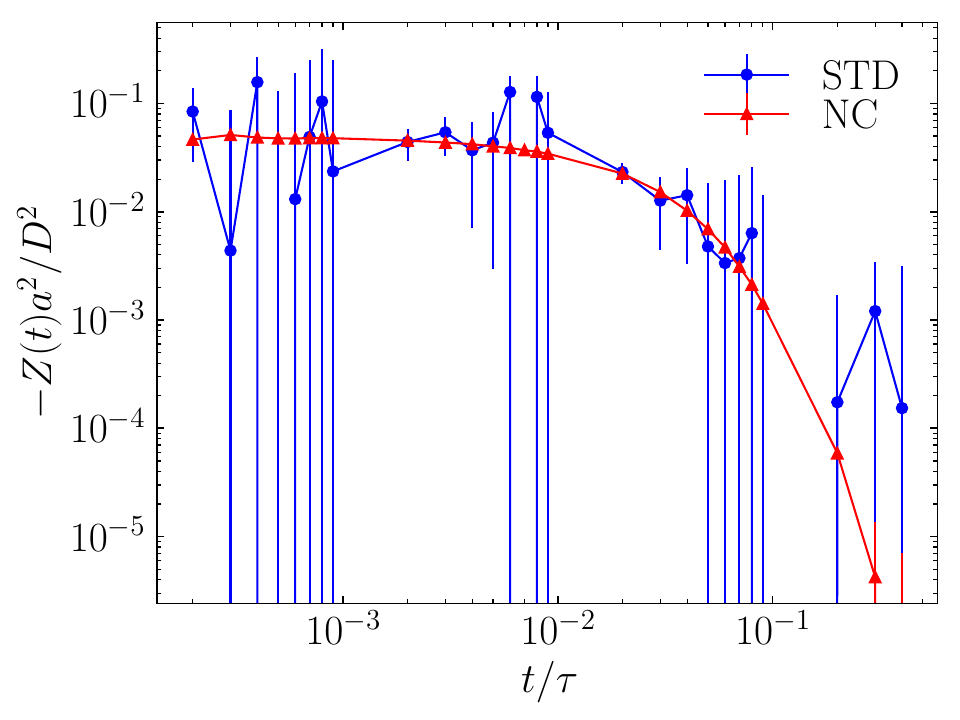}
	\caption{VACF $Z(t)$ for the cosine potential computed using the two different algorithms on double-logarithmic scales for $\Delta U = 0.1 k_\text{B}T$. $Z(t)$ from the standard (STD) method is shown in dots and $Z_\text{NC}(t)$ from the NC algorithm is shown in triangles.}
	\label{fig:VACF_cosinus}
\end{figure}

Finally, we use the algorithm to compute the VACF and visualize the results in~\cref{fig:VACF_cosinus} for the system with $\Delta U = 0.1 k_\text{B} T.$ The figure reveals that for  weakly interacting systems the algorithm indeed becomes increasingly efficient and enables us to extract transport quantities which would otherwise be inaccessible. While the diffusion coefficient exhibits a minimal suppression for lower amplitudes and only minor changes in the MSD emerge, this effect becomes significant in the VACF computation, highlighting the efficiency of the NC algorithm.

\section{Summary and Conclusion} \label{sec:summary}

In this paper we have analyzed the NC algorithm for single Brownian particles in three different external potentials to understand for which systems it can be applied most efficiently. Detailed suggestions for the implementation in Brownian dynamics simulations and especially in Monte Carlo simulations were described, while we highlighted that the implementation is computationally efficient and straightforward. We have demonstrated that the cross-correlation term is negligible in all cases, which significantly simplifies the algorithm. If not, the performance is additionally dependent on the noise originating from the cross-correlation term. Solving the discrete Langevin equation analytically for the harmonic potential model showed that the CC term converges to zero and also that the variance of displacements is not Boltzmann distributed for non-zero simulation time steps but includes a correction term.

We analyzed the performance of the NC algorithm for different systems with varying interaction strengths and found that the performance increases strongly as the interaction becomes weaker.  In contrast, we have highlighted and rationalized that in the case of bounded particles the NC algorithm becomes less efficient than standard simulations. Therefore, the algorithm should be applied to weakly interacting systems, which are notoriously difficult to handle using standard computer simulations. ``Weakly interacting'' can relate to the motion of Brownian particles in weak external potentials, weakly interacting Brownian particles, or dilute suspensions of Brownian particles. In these systems, the correction of the mean-square displacement from free diffusion is minor, whereas this correction is significant for computing the VACF, thus the impact of the NC algorithm is substantial.

This conclusion appears to be independent of details in the interaction potentials and can be applied to both continuous and discontinuous potentials. For simplicity, we have focused in this paper on one-dimensional models of single Brownian particles, but the algorithm is easily extendable to two or more dimensions and to interacting Brownian particles, for which it has originally been developed~\cite{Mandal_2019}. 
Importantly, the algorithm is also not restricted to translational Brownian noise. For example, it should be applicable to reveal complex transport in systems with aspherical particles with rotational diffusion, including the paradigmatic active Brownian particle model~\cite{Breoni_2020}. 
The applicability is not restricted to Brownian noise but can also suppress noise from Monte Carlo simulations. The only requirement is that one can separate the dynamics into an analytically solvable ``free'' motion and complex interactions, which includes for example active Ornstein-Uhlenbeck particles \cite{Fodor_2016} in similar potentials as analyzed in this paper which can induce non-equilibrium transport \cite{Jung_2023}. Thus, the algorithm works only for computer experiments where the pseudo-random noise is known and not for laboratory experiments. 
We therefore expect this algorithm to find various applications in statistical physics to compare with analytical results in weakly interacting systems and in soft matter of passive and active particles to study complex transport.

\section*{Acknowledgements}
We thank Michele Caraglio for helpful discussions. This work has been supported by Austrian Science Fund (FWF) Grant No. I 5257-N.


\begin{thebibliography}{45}%
	\makeatletter
	\providecommand \@ifxundefined [1]{%
	 \@ifx{#1\undefined}
	}%
	\providecommand \@ifnum [1]{%
	 \ifnum #1\expandafter \@firstoftwo
	 \else \expandafter \@secondoftwo
	 \fi
	}%
	\providecommand \@ifx [1]{%
	 \ifx #1\expandafter \@firstoftwo
	 \else \expandafter \@secondoftwo
	 \fi
	}%
	\providecommand \natexlab [1]{#1}%
	\providecommand \enquote  [1]{#1}%
	\providecommand \bibnamefont  [1]{#1}%
	\providecommand \bibfnamefont [1]{#1}%
	\providecommand \citenamefont [1]{#1}%
	\providecommand \href@noop [0]{\@secondoftwo}%
	\providecommand \href[0]{\begingroup \@sanitize@url \@href}%
	\providecommand \@href[1]{\@@startlink{#1}\@@href}%
	\providecommand \@@href[1]{\endgroup#1\@@endlink}%
	\providecommand \@sanitize@url [0]{\catcode `\\12\catcode `\$12\catcode `\&12\catcode `\#12\catcode `\^12\catcode `\_12\catcode `\%12\relax}%
	\providecommand \@@startlink[1]{}%
	\providecommand \@@endlink[0]{}%
	\providecommand \url  [0]{\begingroup\@sanitize@url \@url }%
	\providecommand \@url [1]{\endgroup\@href {#1}{\urlprefix }}%
	\providecommand \urlprefix  [0]{URL }%
	\providecommand \Eprint [0]{\href }%
	\providecommand \doibase [0]{http://dx.doi.org/}%
	\providecommand \selectlanguage [0]{\@gobble}%
	\providecommand \bibinfo  [0]{\@secondoftwo}%
	\providecommand \bibfield  [0]{\@secondoftwo}%
	\providecommand \translation [1]{[#1]}%
	\providecommand \BibitemOpen [0]{}%
	\providecommand \bibitemStop [0]{}%
	\providecommand \bibitemNoStop [0]{.\EOS\space}%
	\providecommand \EOS [0]{\spacefactor3000\relax}%
	\providecommand \BibitemShut  [1]{\csname bibitem#1\endcsname}%
	\let\auto@bib@innerbib\@empty
	\bibitem [{\citenamefont {Einstein}(1905)}]{Einstein_1905}%
	  \BibitemOpen
	  \bibfield  {author} {\bibinfo {author} {\bibfnamefont {A.}~\bibnamefont {Einstein}},\ }\bibfield  {title} {\enquote {\bibinfo {title} {{\"U}ber die von der molekularkinetischen {T}heorie der {W}\"arme geforderte {B}ewegung von in ruhenden {F}l\"ussigkeiten suspendierten {T}eilchen},}\ }\href{\doibase https://doi.org/10.1002/andp.19053220806} {\bibfield  {journal} {\bibinfo  {journal} {Annalen der Physik}\ }\textbf {\bibinfo {volume} {322}},\ \bibinfo {pages} {549} (\bibinfo {year} {1905})}\BibitemShut {NoStop}%
	\bibitem [{\citenamefont {H\"anggi}\ and\ \citenamefont {Marchesoni}(2005)}]{Hanggi_2005}%
	  \BibitemOpen
	  \bibfield  {author} {\bibinfo {author} {\bibfnamefont {P.}~\bibnamefont {H\"anggi}}\ and\ \bibinfo {author} {\bibfnamefont {F.}~\bibnamefont {Marchesoni}},\ }\bibfield  {title} {\enquote {\bibinfo {title} {{Introduction: 100 years of {B}rownian motion}},}\ }\href{\doibase 10.1063/1.1895505} {\bibfield  {journal} {\bibinfo  {journal} {Chaos: An Interdisciplinary Journal of Nonlinear Science}\ }\textbf {\bibinfo {volume} {15}},\ \bibinfo {pages} {026101} (\bibinfo {year} {2005})}\BibitemShut {NoStop}%
	\bibitem [{\citenamefont {Bian}\ \emph {et~al.}(2016)\citenamefont {Bian}, \citenamefont {Kim},\ and\ \citenamefont {Karniadakis}}]{Xin_2016}%
	  \BibitemOpen
	  \bibfield  {author} {\bibinfo {author} {\bibfnamefont {X.}~\bibnamefont {Bian}}, \bibinfo {author} {\bibfnamefont {C.}~\bibnamefont {Kim}}, \ and\ \bibinfo {author} {\bibfnamefont {G.}~\bibnamefont {Karniadakis}},\ }\bibfield  {title} {\enquote {\bibinfo {title} {111 years of {B}rownian motion},}\ }\href{http://dx.doi.org/10.1039/C6SM01153E} {\bibfield  {journal} {\bibinfo  {journal} {Soft Matter}\ }\textbf {\bibinfo {volume} {12}},\ \bibinfo {pages} {6331} (\bibinfo {year} {2016})}\BibitemShut {NoStop}%
	\bibitem [{\citenamefont {Frey}\ and\ \citenamefont {Kroy}(2005)}]{Frey_2005}%
	  \BibitemOpen
	  \bibfield  {author} {\bibinfo {author} {\bibfnamefont {E.}~\bibnamefont {Frey}}\ and\ \bibinfo {author} {\bibfnamefont {K.}~\bibnamefont {Kroy}},\ }\bibfield  {title} {\enquote {\bibinfo {title} {{B}rownian motion: a paradigm of soft matter and biological physics},}\ }\href{\doibase https://doi.org/10.1002/andp.200551701-303} {\bibfield  {journal} {\bibinfo  {journal} {Annalen der Physik}\ }\textbf {\bibinfo {volume} {517}},\ \bibinfo {pages} {20} (\bibinfo {year} {2005})}\BibitemShut {NoStop}%
	\bibitem [{\citenamefont {Bressloff}\ and\ \citenamefont {Newby}(2013)}]{Bressloff_2013}%
	  \BibitemOpen
	  \bibfield  {author} {\bibinfo {author} {\bibfnamefont {P.~C.}\ \bibnamefont {Bressloff}}\ and\ \bibinfo {author} {\bibfnamefont {J.~M.}\ \bibnamefont {Newby}},\ }\bibfield  {title} {\enquote {\bibinfo {title} {Stochastic models of intracellular transport},}\ }\href{\doibase 10.1103/RevModPhys.85.135} {\bibfield  {journal} {\bibinfo  {journal} {Rev. Mod. Phys.}\ }\textbf {\bibinfo {volume} {85}},\ \bibinfo {pages} {135} (\bibinfo {year} {2013})}\BibitemShut {NoStop}%
	\bibitem [{\citenamefont {Enkavi}\ \emph {et~al.}(2019)\citenamefont {Enkavi}, \citenamefont {Javanainen}, \citenamefont {Kulig}, \citenamefont {R\'{o}g},\ and\ \citenamefont {Vattulainen}}]{Enkavi_2019}%
	  \BibitemOpen
	  \bibfield  {author} {\bibinfo {author} {\bibfnamefont {G.}~\bibnamefont {Enkavi}}, \bibinfo {author} {\bibfnamefont {M.}~\bibnamefont {Javanainen}}, \bibinfo {author} {\bibfnamefont {W.}~\bibnamefont {Kulig}}, \bibinfo {author} {\bibfnamefont {T.}~\bibnamefont {R\'{o}g}}, \ and\ \bibinfo {author} {\bibfnamefont {I.}~\bibnamefont {Vattulainen}},\ }\bibfield  {title} {\enquote {\bibinfo {title} {Multiscale simulations of biological membranes: The challenge to understand biological phenomena in a living substance},}\ }\href{https://pubs.acs.org/doi/10.1021/acs.chemrev.8b00538} {\bibfield  {journal} {\bibinfo  {journal} {Chemical Reviews}\ }\textbf {\bibinfo {volume} {119}},\ \bibinfo {pages} {5607–5774} (\bibinfo {year} {2019})}\BibitemShut {NoStop}%
	\bibitem [{\citenamefont {Saxton}\ and\ \citenamefont {Jacobson}(1997)}]{Saxton_1997}%
	  \BibitemOpen
	  \bibfield  {author} {\bibinfo {author} {\bibfnamefont {M.~J.}\ \bibnamefont {Saxton}}\ and\ \bibinfo {author} {\bibfnamefont {K.}~\bibnamefont {Jacobson}},\ }\bibfield  {title} {\enquote {\bibinfo {title} {Single-particle tracking: Applications to membrane dynamics},}\ }\href{\doibase 10.1146/annurev.biophys.26.1.373} {\bibfield  {journal} {\bibinfo  {journal} {Annual Review of Biophysics and Biomolecular Structure}\ }\textbf {\bibinfo {volume} {26}},\ \bibinfo {pages} {373} (\bibinfo {year} {1997})}\BibitemShut {NoStop}%
	\bibitem [{\citenamefont {Franosch}\ \emph {et~al.}(2011)\citenamefont {Franosch}, \citenamefont {Grimm}, \citenamefont {Belushkin}, \citenamefont {Mor}, \citenamefont {Foffi}, \citenamefont {Forr{\'o}},\ and\ \citenamefont {Jeney}}]{Franosch_2011}%
	  \BibitemOpen
	  \bibfield  {author} {\bibinfo {author} {\bibfnamefont {T.}~\bibnamefont {Franosch}}, \bibinfo {author} {\bibfnamefont {M.}~\bibnamefont {Grimm}}, \bibinfo {author} {\bibfnamefont {M.}~\bibnamefont {Belushkin}}, \bibinfo {author} {\bibfnamefont {F.~M.}\ \bibnamefont {Mor}}, \bibinfo {author} {\bibfnamefont {G.}~\bibnamefont {Foffi}}, \bibinfo {author} {\bibfnamefont {L.}~\bibnamefont {Forr{\'o}}}, \ and\ \bibinfo {author} {\bibfnamefont {S.}~\bibnamefont {Jeney}},\ }\bibfield  {title} {\enquote {\bibinfo {title} {Resonances arising from hydrodynamic memory in {B}rownian motion},}\ }\href{\doibase https://doi.org/10.1038/nature10498} {\bibfield  {journal} {\bibinfo  {journal} {Nature}\ }\textbf {\bibinfo {volume} {478}},\ \bibinfo {pages} {85} (\bibinfo {year} {2011})}\BibitemShut {NoStop}%
	\bibitem [{\citenamefont {Michalet}(2010)}]{Xavier_2010}%
	  \BibitemOpen
	  \bibfield  {author} {\bibinfo {author} {\bibfnamefont {X.}~\bibnamefont {Michalet}},\ }\bibfield  {title} {\enquote {\bibinfo {title} {Mean square displacement analysis of single-particle trajectories with localization error: {B}rownian motion in an isotropic medium},}\ }\href{\doibase 10.1103/PhysRevE.82.041914} {\bibfield  {journal} {\bibinfo  {journal} {Phys. Rev. E}\ }\textbf {\bibinfo {volume} {82}},\ \bibinfo {pages} {041914} (\bibinfo {year} {2010})}\BibitemShut {NoStop}%
	\bibitem [{\citenamefont {Alder}\ and\ \citenamefont {Wainwright}(1967)}]{Alder_1967}%
	  \BibitemOpen
	  \bibfield  {author} {\bibinfo {author} {\bibfnamefont {B.~J.}\ \bibnamefont {Alder}}\ and\ \bibinfo {author} {\bibfnamefont {T.~E.}\ \bibnamefont {Wainwright}},\ }\bibfield  {title} {\enquote {\bibinfo {title} {{Velocity Autocorrelations for Hard Spheres}},}\ }\href{\doibase 10.1103/PhysRevLett.18.988} {\bibfield  {journal} {\bibinfo  {journal} {Phys. Rev. Lett.}\ }\textbf {\bibinfo {volume} {18}},\ \bibinfo {pages} {988} (\bibinfo {year} {1967})}\BibitemShut {NoStop}%
	\bibitem [{\citenamefont {Pancorbo}\ \emph {et~al.}(2017)\citenamefont {Pancorbo}, \citenamefont {Rubio},\ and\ \citenamefont {Dominguez-Garcia}}]{Pancorbo_2017}%
	  \BibitemOpen
	  \bibfield  {author} {\bibinfo {author} {\bibfnamefont {M.}~\bibnamefont {Pancorbo}}, \bibinfo {author} {\bibfnamefont {M.~A.}\ \bibnamefont {Rubio}}, \ and\ \bibinfo {author} {\bibfnamefont {P.}~\bibnamefont {Dominguez-Garcia}},\ }\bibfield  {title} {\enquote {\bibinfo {title} {{B}rownian dynamics simulations to explore experimental microsphere diffusion with optical tweezers},}\ }\href{\doibase https://doi.org/10.1016/j.procs.2017.05.231} {\bibfield  {journal} {\bibinfo  {journal} {Procedia Computer Science}\ }\textbf {\bibinfo {volume} {108}},\ \bibinfo {pages} {166} (\bibinfo {year} {2017})},\ \bibinfo {note} {{I}nternational Conference on Computational Science, ICCS 2017, 12-14 June 2017, Zurich, Switzerland}\BibitemShut {NoStop}%
	\bibitem [{\citenamefont {Zembrzycki}\ \emph {et~al.}(2023)\citenamefont {Zembrzycki}, \citenamefont {Pawlowska}, \citenamefont {Pierini},\ and\ \citenamefont {Kowalewski}}]{Zembrzycki_2023}%
	  \BibitemOpen
	  \bibfield  {author} {\bibinfo {author} {\bibfnamefont {K.}~\bibnamefont {Zembrzycki}}, \bibinfo {author} {\bibfnamefont {S.}~\bibnamefont {Pawlowska}}, \bibinfo {author} {\bibfnamefont {F.}~\bibnamefont {Pierini}}, \ and\ \bibinfo {author} {\bibfnamefont {T.~A.}\ \bibnamefont {Kowalewski}},\ }\bibfield  {title} {\enquote {\bibinfo {title} {{B}rownian motion in optical tweezers, a comparison between {M}{D} simulations and experimental data in the ballistic regime},}\ }\href{https://www.mdpi.com/2073-4360/15/3/787} {\bibfield  {journal} {\bibinfo  {journal} {Polymers}\ }\textbf {\bibinfo {volume} {15}},\ \bibinfo {pages} {787} (\bibinfo {year} {2023})}\BibitemShut {NoStop}%
	\bibitem [{\citenamefont {Qi}\ and\ \citenamefont {Schmid}(2017)}]{Qi_2017}%
	  \BibitemOpen
	  \bibfield  {author} {\bibinfo {author} {\bibfnamefont {S.}~\bibnamefont {Qi}}\ and\ \bibinfo {author} {\bibfnamefont {F.}~\bibnamefont {Schmid}},\ }\bibfield  {title} {\enquote {\bibinfo {title} {Dynamic density functional theories for inhomogeneous polymer systems compared to {B}rownian dynamics simulations},}\ }\href{https://pubs.acs.org/doi/10.1021/acs.macromol.7b02017} {\bibfield  {journal} {\bibinfo  {journal} {Macromolecules}\ }\textbf {\bibinfo {volume} {50}},\ \bibinfo {pages} {9831–9845} (\bibinfo {year} {2017})}\BibitemShut {NoStop}%
	\bibitem [{\citenamefont {Dubbeldam}\ \emph {et~al.}(2009)\citenamefont {Dubbeldam}, \citenamefont {Ford}, \citenamefont {Ellis},\ and\ \citenamefont {Snurr}}]{Dubbeldam_2009}%
	  \BibitemOpen
	  \bibfield  {author} {\bibinfo {author} {\bibfnamefont {D.}~\bibnamefont {Dubbeldam}}, \bibinfo {author} {\bibfnamefont {D.~C.}\ \bibnamefont {Ford}}, \bibinfo {author} {\bibfnamefont {D.~E.}\ \bibnamefont {Ellis}}, \ and\ \bibinfo {author} {\bibfnamefont {R.~Q.}\ \bibnamefont {Snurr}},\ }\bibfield  {title} {\enquote {\bibinfo {title} {A new perspective on the order-n algorithm for computing correlation functions},}\ }\href{\doibase 10.1080/08927020902818039} {\bibfield  {journal} {\bibinfo  {journal} {Molecular Simulation}\ }\textbf {\bibinfo {volume} {35}},\ \bibinfo {pages} {1084} (\bibinfo {year} {2009})}\BibitemShut {NoStop}%
	\bibitem [{\citenamefont {Yang}\ \emph {et~al.}(2019)\citenamefont {Yang}, \citenamefont {Shao}, \citenamefont {Zhang}, \citenamefont {Yang},\ and\ \citenamefont {Gao}}]{Yang_2019}%
	  \BibitemOpen
	  \bibfield  {author} {\bibinfo {author} {\bibfnamefont {Y.~I.}\ \bibnamefont {Yang}}, \bibinfo {author} {\bibfnamefont {Q.}~\bibnamefont {Shao}}, \bibinfo {author} {\bibfnamefont {J.}~\bibnamefont {Zhang}}, \bibinfo {author} {\bibfnamefont {L.}~\bibnamefont {Yang}}, \ and\ \bibinfo {author} {\bibfnamefont {Y.~Q.}\ \bibnamefont {Gao}},\ }\bibfield  {title} {\enquote {\bibinfo {title} {{Enhanced sampling in molecular dynamics}},}\ }\href{\doibase 10.1063/1.5109531} {\bibfield  {journal} {\bibinfo  {journal} {The Journal of Chemical Physics}\ }\textbf {\bibinfo {volume} {151}},\ \bibinfo {pages} {070902} (\bibinfo {year} {2019})}\BibitemShut {NoStop}%
	\bibitem [{\citenamefont {Akhmatskaya}\ and\ \citenamefont {Reich}(2008)}]{Akhmatskaya_2008}%
	  \BibitemOpen
	  \bibfield  {author} {\bibinfo {author} {\bibfnamefont {E.}~\bibnamefont {Akhmatskaya}}\ and\ \bibinfo {author} {\bibfnamefont {S.}~\bibnamefont {Reich}},\ }\bibfield  {title} {\enquote {\bibinfo {title} {Gshmc: An efficient method for molecular simulation},}\ }\href{\doibase https://doi.org/10.1016/j.jcp.2008.01.023} {\bibfield  {journal} {\bibinfo  {journal} {Journal of Computational Physics}\ }\textbf {\bibinfo {volume} {227}},\ \bibinfo {pages} {4934} (\bibinfo {year} {2008})}\BibitemShut {NoStop}%
	\bibitem [{\citenamefont {Bernardi}\ \emph {et~al.}(2015)\citenamefont {Bernardi}, \citenamefont {Melo},\ and\ \citenamefont {Schulten}}]{Bernardi_2015}%
	  \BibitemOpen
	  \bibfield  {author} {\bibinfo {author} {\bibfnamefont {R.~C.}\ \bibnamefont {Bernardi}}, \bibinfo {author} {\bibfnamefont {M.~C.}\ \bibnamefont {Melo}}, \ and\ \bibinfo {author} {\bibfnamefont {K.}~\bibnamefont {Schulten}},\ }\bibfield  {title} {\enquote {\bibinfo {title} {Enhanced sampling techniques in molecular dynamics simulations of biological systems},}\ }\href{\doibase https://doi.org/10.1016/j.bbagen.2014.10.019} {\bibfield  {journal} {\bibinfo  {journal} {Biochimica et Biophysica Acta (BBA) - General Subjects}\ }\textbf {\bibinfo {volume} {1850}},\ \bibinfo {pages} {872} (\bibinfo {year} {2015})},\ \bibinfo {note} {recent developments of molecular dynamics}\BibitemShut {NoStop}%
	\bibitem [{\citenamefont {H\'{e}nin}\ \emph {et~al.}(2022)\citenamefont {H\'{e}nin}, \citenamefont {Leli\`{e}vre}, \citenamefont {Shirts}, \citenamefont {Valsson},\ and\ \citenamefont {Delemotte}}]{Henin_2022}%
	  \BibitemOpen
	  \bibfield  {author} {\bibinfo {author} {\bibfnamefont {J.}~\bibnamefont {H\'{e}nin}}, \bibinfo {author} {\bibfnamefont {T.}~\bibnamefont {Leli\`{e}vre}}, \bibinfo {author} {\bibfnamefont {M.}~\bibnamefont {Shirts}}, \bibinfo {author} {\bibfnamefont {O.}~\bibnamefont {Valsson}}, \ and\ \bibinfo {author} {\bibfnamefont {L.}~\bibnamefont {Delemotte}},\ }\bibfield  {title} {\enquote {\bibinfo {title} {Enhanced sampling methods for molecular dynamics simulations},}\ }\href{\doibase 10.33011/livecoms.4.1.1583} {\bibfield  {journal} {\bibinfo  {journal} {Living Journal of Computational Molecular Science}\ }\textbf {\bibinfo {volume} {4}},\ \bibinfo {pages} {1583} (\bibinfo {year} {2022})}\BibitemShut {NoStop}%
	\bibitem [{\citenamefont {Mandal}\ \emph {et~al.}(2019)\citenamefont {Mandal}, \citenamefont {Schrack}, \citenamefont {L\"owen}, \citenamefont {Sperl},\ and\ \citenamefont {Franosch}}]{Mandal_2019}%
	  \BibitemOpen
	  \bibfield  {author} {\bibinfo {author} {\bibfnamefont {S.}~\bibnamefont {Mandal}}, \bibinfo {author} {\bibfnamefont {L.}~\bibnamefont {Schrack}}, \bibinfo {author} {\bibfnamefont {H.}~\bibnamefont {L\"owen}}, \bibinfo {author} {\bibfnamefont {M.}~\bibnamefont {Sperl}}, \ and\ \bibinfo {author} {\bibfnamefont {T.}~\bibnamefont {Franosch}},\ }\bibfield  {title} {\enquote {\bibinfo {title} {{Persistent Anti-Correlations in {B}rownian Dynamics Simulations of Dense Colloidal Suspensions Revealed by Noise Suppression}},}\ }\href{\doibase 10.1103/PhysRevLett.123.168001} {\bibfield  {journal} {\bibinfo  {journal} {Phys. Rev. Lett.}\ }\textbf {\bibinfo {volume} {123}},\ \bibinfo {pages} {168001} (\bibinfo {year} {2019})}\BibitemShut {NoStop}%
	\bibitem [{\citenamefont {Frenkel}(1987)}]{Frenkel_1987}%
	  \BibitemOpen
	  \bibfield  {author} {\bibinfo {author} {\bibfnamefont {D.}~\bibnamefont {Frenkel}},\ }\bibfield  {title} {\enquote {\bibinfo {title} {{Velocity auto-correlation functions in a 2d lattice Lorentz gas: Comparison of theory and computer simulation}},}\ }\href{\doibase https://doi.org/10.1016/0375-9601(87)90482-8} {\bibfield  {journal} {\bibinfo  {journal} {Physics Letters A}\ }\textbf {\bibinfo {volume} {121}},\ \bibinfo {pages} {385} (\bibinfo {year} {1987})}\BibitemShut {NoStop}%
	\bibitem [{\citenamefont {Hanna}\ \emph {et~al.}(1981)\citenamefont {Hanna}, \citenamefont {Hess},\ and\ \citenamefont {Klein}}]{Hanna_1981}%
	  \BibitemOpen
	  \bibfield  {author} {\bibinfo {author} {\bibfnamefont {S.}~\bibnamefont {Hanna}}, \bibinfo {author} {\bibfnamefont {W.}~\bibnamefont {Hess}}, \ and\ \bibinfo {author} {\bibfnamefont {R.}~\bibnamefont {Klein}},\ }\bibfield  {title} {\enquote {\bibinfo {title} {{The velocity autocorrelation function of an overdamped {B}rownian system with hard-core intraction}},}\ }\href{\doibase 10.1088/0305-4470/14/12/004} {\bibfield  {journal} {\bibinfo  {journal} {Journal of Physics A: Mathematical and General}\ }\textbf {\bibinfo {volume} {14}},\ \bibinfo {pages} {L493} (\bibinfo {year} {1981})}\BibitemShut {NoStop}%
	\bibitem [{\citenamefont {Ackerson}\ and\ \citenamefont {Fleishman}(1982)}]{Ackerson_1982}%
	  \BibitemOpen
	  \bibfield  {author} {\bibinfo {author} {\bibfnamefont {B.~J.}\ \bibnamefont {Ackerson}}\ and\ \bibinfo {author} {\bibfnamefont {L.}~\bibnamefont {Fleishman}},\ }\bibfield  {title} {\enquote {\bibinfo {title} {{Correlations for dilute hard core suspensions}},}\ }\href{\doibase 10.1063/1.443251} {\bibfield  {journal} {\bibinfo  {journal} {The Journal of Chemical Physics}\ }\textbf {\bibinfo {volume} {76}},\ \bibinfo {pages} {2675} (\bibinfo {year} {1982})}\BibitemShut {NoStop}%
	\bibitem [{\citenamefont {Felderhof}\ and\ \citenamefont {Jones}(1983)}]{Felderhof_1983}%
	  \BibitemOpen
	  \bibfield  {author} {\bibinfo {author} {\bibfnamefont {B.}~\bibnamefont {Felderhof}}\ and\ \bibinfo {author} {\bibfnamefont {R.}~\bibnamefont {Jones}},\ }\bibfield  {title} {\enquote {\bibinfo {title} {Diffusion in hard sphere suspensions},}\ }\href{\doibase https://doi.org/10.1016/0378-4371(83)90084-5} {\bibfield  {journal} {\bibinfo  {journal} {Physica A: Statistical Mechanics and its Applications}\ }\textbf {\bibinfo {volume} {122}},\ \bibinfo {pages} {89} (\bibinfo {year} {1983})}\BibitemShut {NoStop}%
	\bibitem [{\citenamefont {Lowe}\ \emph {et~al.}(1995)\citenamefont {Lowe}, \citenamefont {Frenkel},\ and\ \citenamefont {Masters}}]{Lowe_1995}%
	  \BibitemOpen
	  \bibfield  {author} {\bibinfo {author} {\bibfnamefont {C.}~\bibnamefont {Lowe}}, \bibinfo {author} {\bibfnamefont {D.}~\bibnamefont {Frenkel}}, \ and\ \bibinfo {author} {\bibfnamefont {A.}~\bibnamefont {Masters}},\ }\bibfield  {title} {\enquote {\bibinfo {title} {Long-time tails in angular momentum correlations - response},}\ }\href{https://pubs.aip.org/aip/jcp/article/103/4/1582/479373/Long-time-tails-in-angular-momentum-correlations} {\bibfield  {journal} {\bibinfo  {journal} {The Journal of Chemical Physics}\ }\textbf {\bibinfo {volume} {104}},\ \bibinfo {pages} {1582–1587} (\bibinfo {year} {1995})}\BibitemShut {NoStop}%
	\bibitem [{\citenamefont {{\"O}ttinger}(1994)}]{Oettinger_1994}%
	  \BibitemOpen
	  \bibfield  {author} {\bibinfo {author} {\bibfnamefont {H.~C.}\ \bibnamefont {{\"O}ttinger}},\ }\bibfield  {title} {\enquote {\bibinfo {title} {Variance reduced brownian dynamics simulations},}\ }\href{\doibase 10.1021/ma00090a041} {\bibfield  {journal} {\bibinfo  {journal} {Macromolecules}\ }\textbf {\bibinfo {volume} {27}},\ \bibinfo {pages} {3415} (\bibinfo {year} {1994})}\BibitemShut {NoStop}%
	\bibitem [{\citenamefont {Melchior}\ and\ \citenamefont {{\"O}ttinger}(1996)}]{Melchior_1996}%
	  \BibitemOpen
	  \bibfield  {author} {\bibinfo {author} {\bibfnamefont {M.}~\bibnamefont {Melchior}}\ and\ \bibinfo {author} {\bibfnamefont {H.~C.}\ \bibnamefont {{\"O}ttinger}},\ }\bibfield  {title} {\enquote {\bibinfo {title} {{Variance reduced simulations of polymer dynamics}},}\ }\href{\doibase 10.1063/1.472186} {\bibfield  {journal} {\bibinfo  {journal} {The Journal of Chemical Physics}\ }\textbf {\bibinfo {volume} {105}},\ \bibinfo {pages} {3316} (\bibinfo {year} {1996})}\BibitemShut {NoStop}%
	\bibitem [{\citenamefont {Evans}\ and\ \citenamefont {Morriss}(2008)}]{Evans_2008}%
	  \BibitemOpen
	  \bibfield  {author} {\bibinfo {author} {\bibfnamefont {D.~J.}\ \bibnamefont {Evans}}\ and\ \bibinfo {author} {\bibfnamefont {G.}~\bibnamefont {Morriss}},\ }\href{\doibase 10.1017/CBO9780511535307} {\emph {\bibinfo {title} {Statistical Mechanics of Nonequilibrium Liquids}}},\ \bibinfo {edition} {2nd}\ ed.\ (\bibinfo  {publisher} {Cambridge University Press},\ \bibinfo {year} {2008})\BibitemShut {NoStop}%
	\bibitem [{\citenamefont {Brinkman}(1956)}]{Brinkman_1956}%
	  \BibitemOpen
	  \bibfield  {author} {\bibinfo {author} {\bibfnamefont {H.}~\bibnamefont {Brinkman}},\ }\bibfield  {title} {\enquote {\bibinfo {title} {{B}rownian motion in a field of force and the diffusion theory of chemical reactions},}\ }\href{\doibase https://doi.org/10.1016/S0031-8914(56)80006-2} {\bibfield  {journal} {\bibinfo  {journal} {Physica}\ }\textbf {\bibinfo {volume} {22}},\ \bibinfo {pages} {29} (\bibinfo {year} {1956})}\BibitemShut {NoStop}%
	\bibitem [{\citenamefont {M{\"o}rsch}\ \emph {et~al.}(1979)\citenamefont {M{\"o}rsch}, \citenamefont {Risken},\ and\ \citenamefont {Vollmer}}]{Moersch_1979}%
	  \BibitemOpen
	  \bibfield  {author} {\bibinfo {author} {\bibfnamefont {M.}~\bibnamefont {M{\"o}rsch}}, \bibinfo {author} {\bibfnamefont {H.}~\bibnamefont {Risken}}, \ and\ \bibinfo {author} {\bibfnamefont {H.~D.}\ \bibnamefont {Vollmer}},\ }\bibfield  {title} {\enquote {\bibinfo {title} {{One-dimensional diffusion in soluble model potentials}},}\ }\href{\doibase https://doi.org/10.1007/BF01320120} {\bibfield  {journal} {\bibinfo  {journal} {Zeitschrift f{\"u}r Physik B Condensed Matter}\ }\textbf {\bibinfo {volume} {32}},\ \bibinfo {pages} {245} (\bibinfo {year} {1979})}\BibitemShut {NoStop}%
	\bibitem [{\citenamefont {Capellmann}\ \emph {et~al.}(2018)\citenamefont {Capellmann}, \citenamefont {Khisameeva}, \citenamefont {Platten},\ and\ \citenamefont {Egelhaaf}}]{Capellmann_2018}%
	  \BibitemOpen
	  \bibfield  {author} {\bibinfo {author} {\bibfnamefont {R.~F.}\ \bibnamefont {Capellmann}}, \bibinfo {author} {\bibfnamefont {A.}~\bibnamefont {Khisameeva}}, \bibinfo {author} {\bibfnamefont {F.}~\bibnamefont {Platten}}, \ and\ \bibinfo {author} {\bibfnamefont {S.~U.}\ \bibnamefont {Egelhaaf}},\ }\bibfield  {title} {\enquote {\bibinfo {title} {{Dense colloidal mixtures in an external sinusoidal potential}},}\ }\href{\doibase 10.1063/1.5013007} {\bibfield  {journal} {\bibinfo  {journal} {The Journal of Chemical Physics}\ }\textbf {\bibinfo {volume} {148}},\ \bibinfo {pages} {114903} (\bibinfo {year} {2018})}\BibitemShut {NoStop}%
	\bibitem [{\citenamefont {Ambegaokar}\ and\ \citenamefont {Halperin}(1969)}]{Ambegaokar_1969}%
	  \BibitemOpen
	  \bibfield  {author} {\bibinfo {author} {\bibfnamefont {V.}~\bibnamefont {Ambegaokar}}\ and\ \bibinfo {author} {\bibfnamefont {B.~I.}\ \bibnamefont {Halperin}},\ }\bibfield  {title} {\enquote {\bibinfo {title} {Voltage due to thermal noise in the dc {J}osephson effect},}\ }\href{\doibase 10.1103/PhysRevLett.22.1364} {\bibfield  {journal} {\bibinfo  {journal} {Phys. Rev. Lett.}\ }\textbf {\bibinfo {volume} {22}},\ \bibinfo {pages} {1364} (\bibinfo {year} {1969})}\BibitemShut {NoStop}%
	\bibitem [{\citenamefont {Weiner}\ and\ \citenamefont {Forman}(1974)}]{Weiner_1974}%
	  \BibitemOpen
	  \bibfield  {author} {\bibinfo {author} {\bibfnamefont {J.~H.}\ \bibnamefont {Weiner}}\ and\ \bibinfo {author} {\bibfnamefont {R.~E.}\ \bibnamefont {Forman}},\ }\bibfield  {title} {\enquote {\bibinfo {title} {Rate theory for solids. {IV}. classical {B}rownian-motion model},}\ }\href{\doibase 10.1103/PhysRevB.10.315} {\bibfield  {journal} {\bibinfo  {journal} {Phys. Rev. B}\ }\textbf {\bibinfo {volume} {10}},\ \bibinfo {pages} {315} (\bibinfo {year} {1974})}\BibitemShut {NoStop}%
	\bibitem [{\citenamefont {Sanchez-Palencia}(2004)}]{Sanchez-Palencia_2004}%
	  \BibitemOpen
	  \bibfield  {author} {\bibinfo {author} {\bibfnamefont {L.}~\bibnamefont {Sanchez-Palencia}},\ }\bibfield  {title} {\enquote {\bibinfo {title} {Directed transport of {B}rownian particles in a double symmetric potential},}\ }\href{\doibase 10.1103/PhysRevE.70.011102} {\bibfield  {journal} {\bibinfo  {journal} {Phys. Rev. E}\ }\textbf {\bibinfo {volume} {70}},\ \bibinfo {pages} {011102} (\bibinfo {year} {2004})}\BibitemShut {NoStop}%
	\bibitem [{\citenamefont {Jack}\ and\ \citenamefont {Deaker}(2022)}]{Jack_2022}%
	  \BibitemOpen
	  \bibfield  {author} {\bibinfo {author} {\bibfnamefont {M.~W.}\ \bibnamefont {Jack}}\ and\ \bibinfo {author} {\bibfnamefont {A.}~\bibnamefont {Deaker}},\ }\bibfield  {title} {\enquote {\bibinfo {title} {Nonequilibrium master equation for interacting {B}rownian particles in a deep-well periodic potential},}\ }\href{\doibase 10.1103/PhysRevE.105.054150} {\bibfield  {journal} {\bibinfo  {journal} {Phys. Rev. E}\ }\textbf {\bibinfo {volume} {105}},\ \bibinfo {pages} {054150} (\bibinfo {year} {2022})}\BibitemShut {NoStop}%
	\bibitem [{\citenamefont {Franosch}\ \emph {et~al.}(2010)\citenamefont {Franosch}, \citenamefont {H{\"o}fling}, \citenamefont {Bauer},\ and\ \citenamefont {Frey}}]{Franosch_2010}%
	  \BibitemOpen
	  \bibfield  {author} {\bibinfo {author} {\bibfnamefont {T.}~\bibnamefont {Franosch}}, \bibinfo {author} {\bibfnamefont {F.}~\bibnamefont {H{\"o}fling}}, \bibinfo {author} {\bibfnamefont {T.}~\bibnamefont {Bauer}}, \ and\ \bibinfo {author} {\bibfnamefont {E.}~\bibnamefont {Frey}},\ }\bibfield  {title} {\enquote {\bibinfo {title} {Persistent memory for a {B}rownian walker in a random array of obstacles},}\ }\href{\doibase https://doi.org/10.1016/j.chemphys.2010.04.023} {\bibfield  {journal} {\bibinfo  {journal} {Chemical Physics}\ }\textbf {\bibinfo {volume} {375}},\ \bibinfo {pages} {540} (\bibinfo {year} {2010})},\ \bibinfo {note} {stochastic processes in Physics and Chemistry (in honor of Peter H{\"a}nggi)}\BibitemShut {NoStop}%
	\bibitem [{\citenamefont {Burrage}\ \emph {et~al.}(2000)\citenamefont {Burrage}, \citenamefont {Burrage},\ and\ \citenamefont {Mitsui}}]{Burrage_2000}%
	  \BibitemOpen
	  \bibfield  {author} {\bibinfo {author} {\bibfnamefont {K.}~\bibnamefont {Burrage}}, \bibinfo {author} {\bibfnamefont {P.}~\bibnamefont {Burrage}}, \ and\ \bibinfo {author} {\bibfnamefont {T.}~\bibnamefont {Mitsui}},\ }\bibfield  {title} {\enquote {\bibinfo {title} {Numerical solutions of stochastic differential equations – implementation and stability issues},}\ }\href{\doibase https://doi.org/10.1016/S0377-0427(00)00467-2} {\bibfield  {journal} {\bibinfo  {journal} {Journal of Computational and Applied Mathematics}\ }\textbf {\bibinfo {volume} {125}},\ \bibinfo {pages} {171} (\bibinfo {year} {2000})},\ \bibinfo {note} {numerical Analysis 2000. Vol. VI: Ordinary Differential Equations and Integral Equations}\BibitemShut {NoStop}%
	\bibitem [{\citenamefont {Maruyama}(1955)}]{Maruyama_1955}%
	  \BibitemOpen
	  \bibfield  {author} {\bibinfo {author} {\bibfnamefont {G.}~\bibnamefont {Maruyama}},\ }\bibfield  {title} {\enquote {\bibinfo {title} {{Continuous {M}arkov processes and stochastic equations}},}\ }\href{\doibase https://doi.org/10.1007/BF02846028} {\bibfield  {journal} {\bibinfo  {journal} {Rendiconti del Circolo Matematico di Palermo}\ }\textbf {\bibinfo {volume} {4}},\ \bibinfo {pages} {48} (\bibinfo {year} {1955})}\BibitemShut {NoStop}%
	\bibitem [{\citenamefont {Ermak}\ and\ \citenamefont {McCammon}(1978)}]{Ermak_1978}%
	  \BibitemOpen
	  \bibfield  {author} {\bibinfo {author} {\bibfnamefont {D.~L.}\ \bibnamefont {Ermak}}\ and\ \bibinfo {author} {\bibfnamefont {J.~A.}\ \bibnamefont {McCammon}},\ }\bibfield  {title} {\enquote {\bibinfo {title} {{{B}rownian dynamics with hydrodynamic interactions}},}\ }\href{\doibase 10.1063/1.436761} {\bibfield  {journal} {\bibinfo  {journal} {The Journal of Chemical Physics}\ }\textbf {\bibinfo {volume} {69}},\ \bibinfo {pages} {1352} (\bibinfo {year} {1978})}\BibitemShut {NoStop}%
	\bibitem [{\citenamefont {Frenkel}\ and\ \citenamefont {Smit}(2002)}]{Frenkel_2002}%
	  \BibitemOpen
	  \bibfield  {author} {\bibinfo {author} {\bibfnamefont {D.}~\bibnamefont {Frenkel}}\ and\ \bibinfo {author} {\bibfnamefont {B.}~\bibnamefont {Smit}},\ }\bibfield  {title} {\enquote {\bibinfo {title} {Chapter 4 - molecular dynamics simulations},}\ }in\ \href{\doibase https://doi.org/10.1016/B978-012267351-1/50006-7} {\emph {\bibinfo {booktitle} {Understanding Molecular Simulation (Second Edition)}}},\ \bibinfo {editor} {edited by\ \bibinfo {editor} {\bibfnamefont {D.}~\bibnamefont {Frenkel}}\ and\ \bibinfo {editor} {\bibfnamefont {B.}~\bibnamefont {Smit}}}\ (\bibinfo  {publisher} {Academic Press},\ \bibinfo {address} {San Diego},\ \bibinfo {year} {2002})\ \bibinfo {edition} {second edition}\ ed.,\ pp.\ \bibinfo {pages} {63--107}\BibitemShut {NoStop}%
	\bibitem [{\citenamefont {Allen}\ and\ \citenamefont {Tildesley}(2017)}]{Allen_2017}%
	  \BibitemOpen
	  \bibfield  {author} {\bibinfo {author} {\bibfnamefont {M.~P.}\ \bibnamefont {Allen}}\ and\ \bibinfo {author} {\bibfnamefont {D.~J.}\ \bibnamefont {Tildesley}},\ }\href{\doibase 10.1093/oso/9780198803195.001.0001} {\emph {\bibinfo {title} {{Computer Simulation of Liquids}}}}\ (\bibinfo  {publisher} {Oxford University Press},\ \bibinfo {year} {2017})\BibitemShut {NoStop}%
	\bibitem [{\citenamefont {Metropolis}\ \emph {et~al.}(1953)\citenamefont {Metropolis}, \citenamefont {Rosenbluth}, \citenamefont {Rosenbluth}, \citenamefont {Teller},\ and\ \citenamefont {Teller}}]{Metropolis_1953}%
	  \BibitemOpen
	  \bibfield  {author} {\bibinfo {author} {\bibfnamefont {N.}~\bibnamefont {Metropolis}}, \bibinfo {author} {\bibfnamefont {A.~W.}\ \bibnamefont {Rosenbluth}}, \bibinfo {author} {\bibfnamefont {M.~N.}\ \bibnamefont {Rosenbluth}}, \bibinfo {author} {\bibfnamefont {A.~H.}\ \bibnamefont {Teller}}, \ and\ \bibinfo {author} {\bibfnamefont {E.}~\bibnamefont {Teller}},\ }\bibfield  {title} {\enquote {\bibinfo {title} {{Equation of State Calculations by Fast Computing Machines}},}\ }\href{\doibase 10.1063/1.1699114} {\bibfield  {journal} {\bibinfo  {journal} {The Journal of Chemical Physics}\ }\textbf {\bibinfo {volume} {21}},\ \bibinfo {pages} {1087} (\bibinfo {year} {1953})}\BibitemShut {NoStop}%
	\bibitem [{\citenamefont {Hansen}\ and\ \citenamefont {McDonald}(1990)}]{Hansen_2007}%
	  \BibitemOpen
	  \bibfield  {author} {\bibinfo {author} {\bibfnamefont {J.~P.}\ \bibnamefont {Hansen}}\ and\ \bibinfo {author} {\bibfnamefont {I.}~\bibnamefont {McDonald}},\ }\href@noop {} {\emph {\bibinfo {title} {{Theory of Simple Liquids}}}}\ (\bibinfo  {publisher} {Academic},\ \bibinfo {address} {London},\ \bibinfo {year} {1990})\BibitemShut {NoStop}%
	\bibitem [{\citenamefont {Breoni}\ \emph {et~al.}(2020)\citenamefont {Breoni}, \citenamefont {Schmiedeberg},\ and\ \citenamefont {L{\"o}wen}}]{Breoni_2020}%
	  \BibitemOpen
	  \bibfield  {author} {\bibinfo {author} {\bibfnamefont {D.}~\bibnamefont {Breoni}}, \bibinfo {author} {\bibfnamefont {M.}~\bibnamefont {Schmiedeberg}}, \ and\ \bibinfo {author} {\bibfnamefont {H.}~\bibnamefont {L{\"o}wen}},\ }\bibfield  {title} {\enquote {\bibinfo {title} {Active {B}rownian and inertial particles in disordered environments: Short-time expansion of the mean-square displacement},}\ }\href{\doibase 10.1103/PhysRevE.102.062604} {\bibfield  {journal} {\bibinfo  {journal} {Phys. Rev. E}\ }\textbf {\bibinfo {volume} {102}},\ \bibinfo {pages} {062604} (\bibinfo {year} {2020})}\BibitemShut {NoStop}%
	\bibitem [{\citenamefont {Fodor}\ \emph {et~al.}(2016)\citenamefont {Fodor}, \citenamefont {Nardini}, \citenamefont {Cates}, \citenamefont {Tailleur}, \citenamefont {Visco},\ and\ \citenamefont {van Wijland}}]{Fodor_2016}%
	  \BibitemOpen
	  \bibfield  {author} {\bibinfo {author} {\bibfnamefont {E.}~\bibnamefont {Fodor}}, \bibinfo {author} {\bibfnamefont {C.}~\bibnamefont {Nardini}}, \bibinfo {author} {\bibfnamefont {M.~E.}\ \bibnamefont {Cates}}, \bibinfo {author} {\bibfnamefont {J.}~\bibnamefont {Tailleur}}, \bibinfo {author} {\bibfnamefont {P.}~\bibnamefont {Visco}}, \ and\ \bibinfo {author} {\bibfnamefont {F.}~\bibnamefont {van Wijland}},\ }\bibfield  {title} {\enquote {\bibinfo {title} {How far from equilibrium is active matter?}}\ }\href{\doibase 10.1103/PhysRevLett.117.038103} {\bibfield  {journal} {\bibinfo  {journal} {Phys. Rev. Lett.}\ }\textbf {\bibinfo {volume} {117}},\ \bibinfo {pages} {038103} (\bibinfo {year} {2016})}\BibitemShut {NoStop}%
	\bibitem [{\citenamefont {{Jung}}(2023)}]{Jung_2023}%
	  \BibitemOpen
	  \bibfield  {author} {\bibinfo {author} {\bibfnamefont {G.}~\bibnamefont {{Jung}}},\ }\bibfield  {title} {\enquote {\bibinfo {title} {{Mobility, response and transport in non-equilibrium coarse-grained models}},}\ }\href{\doibase 10.48550/arXiv.2310.03565} {\bibfield  {journal} {\bibinfo  {journal} {arXiv e-prints}\ ,\ \bibinfo {eid} {arXiv:2310.03565}} (\bibinfo {year} {2023})}\BibitemShut {NoStop}%
	\end{thebibliography}

%

\end{document}